\documentclass[useAMS,usenatbib]{mn2e}
\usepackage{graphicx}
\usepackage{epstopdf}

\title[Cosmology and Cluster Halo Scaling Relations]{Cosmology and Cluster Halo Scaling Relations}
\author[Araya-Melo et al.]{\parbox{\textwidth}{Pablo A. Araya-Melo$^{1,2,3}$, Rien van de Weygaert$^1$ and Bernard J.T. Jones$^1$}
\vspace*{4pt}\\
$^1$Kapteyn Astronomical Institute, University of Groningen, P.O.
Box 800, 9700 AV Groningen, The Netherlands\\
$^2$Korea Institute for Advanced Study, Dongdaemon-gu, Seoul 130-722, Korea\\
$^3$Jacobs University Bremen, Campus Ring 1, 28759 Bremen, Germany}

\begin{document}

\date{}

\pagerange{\pageref{firstpage}--\pageref{lastpage}} \pubyear{2008}

\maketitle

\label{firstpage}

\begin{abstract}
We explore the effects of dark matter and dark energy on the dynamical scaling 
properties of galaxy clusters. We investigate the cluster Faber-Jackson (FJ), 
Kormendy and Fundamental Plane (FP) relations between the mass, radius and 
velocity dispersion of cluster size halos in cosmological $N$-body simulations.
The simulations span a wide range of cosmological parameters, representing 
open, flat and closed Universes. 

Independently of the cosmology, we find that the simulated clusters are close 
to a perfect virial state and do indeed define a Fundamental Plane.  The fitted
parameters of the FJ, Kormendy and FP relationships do not show any significant
dependence on $\Omega_m$ and/or $\Omega_{\Lambda}$.  The one outstanding effect 
is the influence of $\Omega_{m}$ on the thickness of the Fundamental Plane. 

Following the time evolution of our models, we find slight changes of FJ and 
Kormendy parameters in high $\Omega_m$ universe, along with a slight decrease 
of FP fitting parameters.  We also see an initial increase of the FP thickness 
followed by a convergence to a nearly constant value.  The epoch of convergence
is later for higher values of $\Omega_m$ while the thickness remains constant 
in the low $\Omega_m$ $\Lambda$-models.  We also find a continuous increase of 
the FP thickness in the Standard CDM (SCDM) cosmology. There is no evidence 
that these differences are due to the different power spectrum slope at cluster 
scales. 

From the point of view of the FP, there is little difference between clusters 
that quietly accreted their mass and those that underwent massive mergers. The 
principal effect of strong mergers is to change significantly the ratio of the 
half-mass radius $r_{half}$ to the harmonic mean radius $r_h$.  
\end{abstract}


\section{Introduction}
\label{sec5:intro}
Recent observations of distant supernovae \citep{riess98,perlmutter99} suggest 
that we are living in a flat, accelerated Universe with a low matter density.  
This accelerated expansion has established the possibility of a dark energy 
component which behaves like Einstein's cosmological constant $\Lambda$. A 
positive cosmological constant resolves the apparent conflict suggested by the 
old age of globular cluster stars and the estimated value \citep{spergel03,
spergel07} appears sufficient to yield a flat geometry of our Universe.

The role of $\Lambda$  in the process of structure formation is not yet fully 
understood. Although its influence can be seen when looking at the global 
evolution of the Universe, its role in the dynamical evolution of cosmic 
structures is not clear.  The most direct impact of $\Lambda$ comes from its 
influence on the amplitude of the primordial perturbation power spectrum; there
is also an influence from the change in the cosmic and dynamical time scales. 
The direct dynamical influence is probably minor: we do know that in the linear
regime it accounts for a mere $\sim1/70$th of the influence of matter 
perturbation \citep{lahav91}.
 
Most viable theories of cosmic structure formation involve hierarchical 
clustering. Small structures form first and they merge to give birth to bigger 
ones. The rate and history of this process is highly dependent on the amount of
(dark) matter present in the Universe. In Universes with a low $\Omega_{m}$, 
structure formation ceases at much early times than that in cosmologies with 
high density values.

Within this hierarchical process, clusters of galaxies are the most massive and
most recently formed structures in the Universe. Their collapse time is 
comparable to the age of the Universe. This makes them important probes for the
study of cosmic structure formation and evolution. The hierarchical clustering 
history from which galaxy clusters emerge involves a highly complex process of 
merging, accretion and virialization. In this paper we investigate in how far 
we can get insight into this history on the basis of the internal properties of
the clusters. This involves characteristics like their mass and mass 
distribution, their size and their kinetic and gravitational potential energy. 
In particular, we are keen to learn whether these do show any possible trace of
a cosmological constant.

One particular profound manifestation of the virial state of cosmic objects is 
via scaling relations that connect various structural properties.  Scaling 
relations of collapsed and virialized objects relate two or three fundamental 
characteristics. The first involves a quantity measuring the amount of mass 
$M$, often expressed in terms of the amount of light $L$ emitted by the object.
The second quantity involves the size of the object, while the third one 
quantifies its dynamical state. For a virialized halo with mass $M$, size $R$ 
and velocity dispersion $\sigma_{v}=\langle v^{2}\rangle^{1/2}$, the implied 
scaling relation is 
\begin{equation}
\log{M}\,=\,2\log{\sigma_{v}}\,+\,\log{R}\,+\,\epsilon_{M}\,,
\label{eq:fp_virial}
\end{equation}
where $\epsilon_{M}$ is a constant that reflects the internal dynamics of the 
system.  ($\epsilon_{M}$ is determined by issues such as the isotropy if the 
cluster velocity dispersion, its shape and any substructure).  

Systems having similar values of this constant would be expected to form a 
two-parameter family of objects: observationally this manifests itself as the 
``Fundamental Plane''.  Objects lying on the same plane might be expected to 
have similar formation histories and, conversely, the nature of the Fundamental
Plane is a clue to the underlying formation mechanism. 

The scaling relations are of great importance for a variety of reasons.  First 
of all, they inform us about the dynamical state of the objects and must be a 
profound reflection of the galaxy formation process \citep{robertson06}. Also,
they have turned out to be of substantial practical importance. Because they 
relate an intrinsic distance independent quantity like velocity dispersion to a
distance dependent one like $L_{e}$, they can be used as cosmological distance 
indicators.

\subsection{Observed relationships}
\subsubsection{Galaxies}
Since the mid 70s, we know that the observed properties of elliptical galaxies 
follow scaling relations. The Faber-Jackson relation \citep{fj76} relates the 
luminosity $L$ and the velocity dispersion $\sigma$ of an elliptical galaxy. 
The Tully-Fisher relation \citep{tully77} is the equivalent for spiral 
galaxies. A different, though related, scaling is that between the effective 
radius $r_{e}$ and the luminosity $L$ of the galaxy. This is known as the 
Kormendy relation \citep{kormendy77}. These two relations turned out to be 
manifestations of a deeper scaling relation between three fundamental 
characteristics, which became known as the Fundamental Plane 
\citep{djorgovski87, dressler87}. 

The Fundamental Plane is generally expressed as a relationship between three 
parameters, though there is no consensus as to which three should best be used,
nor precisely how to define them.  This makes detailed comparisons somewhat 
difficult. Some authors use the set ($\log R, \log \sigma, \log I$), $I$ being 
the luminosity in some spectral band within some radius $R$, while others use
the set ($\log R, \log \sigma, \mu$), $\mu$ being the mean surface brightness 
within that radius.  Comparisons are further complicated by the fact that there
appear to be manifest residual luminosity dependences in the fits, as reported 
in a recent study of the SDSS by \cite{Nigoche-Netro}.

Care is needed when interpreting these observed relationships.  Observed data 
generally refers to luminosity rather than mass, and the radius that is used 
generally refers to some fiducial radius such as the half-light radius or some 
radius based on profile fitting.  Often, the half-light radius, $R_e$ as 
determined from a fit to a de Vaucouleurs profile is used.  

This situation has been improved somewhat by the gravitational lensing study of
\cite{bolton07}.  These authors presented a new formulation of the FP using 
lensing data to replace surface brightness with surface mass density.  They 
also present an interesting alternative, which they refer to as the ``Mass 
Plane'' (MP), in which they find the dependence of $\log(R_e)$ on 
$\log(\sigma_{e2})$ and surface mass density $\Sigma_{e2}$ within a radius 
$R_e/2$.  Using surface mass density $\Sigma_{e2}$ within a radius $R_e/2$ in 
place of surface brightness $I_e$ removes one of the assumptions about the 
relationship between mass and light.  

\subsubsection{Galaxy Clusters}
Much recent galaxy cluster work on the Fundamental Plane has focussed on the 
differences between the Fundamental Planes of the clusters as defined by their 
member galaxies (see for example \cite{onofrio08} and references therein).

Galaxy cluster scaling relations were discovered by \cite{schaeffer93} who 
studied a sample of 16 galaxy clusters, concluding that these systems also 
populate a Fundamental Plane.  \cite{adami98} used the ESO Nearby Abell Cluster
Survey (ENACS) to study the existence of a Fundamental Plane for rich galaxy 
clusters, finding that it is significantly different from that for elliptical 
galaxies. \cite{marmo04} using data from the WINGS cluster survey found that 
the difference is largely a simple shift in the relative positions of the 
planes.

The largely unknown relationship between mass and light frustrates a direct 
comparison with the results of N-Body investigations.

\subsection{Numerical investigations}
Later, \cite{lanzoni04} addressed the question using N-Body simulations for 
high mass halos, which are thought to host clusters of galaxies. On the basis 
of 13 simulated massive dark matter halos in a $\Lambda$CDM cosmology they 
found that the dark matter halos follow the FJ, Kormendy and FP-like relations. 

\begin {table*}
  \begin {center}
    \begin {tabular}{||c|cccccccc|}
      \hline
      \hline
      Model     & $\Omega_{m}$ & $\Omega_{\Lambda}$ & $\Omega_{k}$ & Age & 
      $m_{dm} $ & $m_{cut}$    & $\Delta_{vir,b}$   & $\Delta_{vir,c}$ \\ 
      \hline
      \hline
      SCDM           & 1.0 & 0 & 0   & 9.31  & 13.23 & 1323 & 177.65 & 177.65\\
      OCDM01         & 0.1 & 0 & 0.9 & 12.55 & 1.32  & 132  & 978.83 & 97.88 \\
      OCDM03         & 0.3 & 0 & 0.7 & 11.30 & 3.97  & 397  & 402.34 & 120.70\\
      OCDM05         & 0.5 & 0 & 0.5 & 10.53 & 6.62  & 662  & 278.10 & 139.05\\
      \hline
      \hline
      $\Lambda$CDMO1 & 0.1 & 0.5 & 0.4 & 14.65  & 1.32 & 132 & 838.30 & 83.83\\
      $\Lambda$CDMO2 & 0.1 & 0.7 & 0.2 & 15.96 & 1.32 & 132 & 778.30 & 77.83\\
      $\Lambda$CDMF1 & 0.1 & 0.9 & 0   & 17.85 & 1.32 & 132 & 715.12 & 71.51\\
      $\Lambda$CDMO3 & 0.3 & 0.5 & 0.2 & 12.70 & 3.97 & 397 & 358.21 & 107.46\\
      $\Lambda$CDMF2 & 0.3 & 0.7 & 0   & 13.47 & 3.97 & 397 & 339.78 & 101.93\\
      $\Lambda$CDMC1 & 0.3 & 0.9 & -0.2& 14.44 & 3.97 & 397 & 320.79 & 96.237\\
      $\Lambda$CDMF3 & 0.5 & 0.5 & 0   & 11.61 & 6.62 & 662 & 252.38 & 126.19\\
      $\Lambda$CDMC2 & 0.5 & 0.7 & -0.2 & 12.17 & 6.62 & 662& 241.74 & 120.87\\
      $\Lambda$CDMC3 & 0.5 & 0.9 & -0.4& 12.84 & 6.62 & 6622 & 30.85 & 115.43\\
      \hline
      \hline
    \end {tabular}
    \caption{\small{Cosmological parameters for the runs. The first column 
gives the identification of the runs, and the following columns give the 
present matter density parameter, the density parameter associated with the 
cosmological constant, $\Omega_k\,=\,1-\Omega_m-\Omega_\Lambda$ quantifies 
the curvature of the Universe, the age of the Universe in Gyr since the Big 
Bang, the mass per particle in units of 10$^{10}h^{-1}$M$_{\odot}$, the mass cut
of the groups given by HOP in units of 10$^{10}h^{-1}$M$_{\odot}$, the value of 
the (over)density needed to have virialized objects with respect to the 
background density, and similarly, but now with respect to the critical 
density.}}
\label{table5:paramsim}
  \end {center}
\end {table*}

In hierarchical scenarios of structure formation halos build up by subsequent 
merging of smaller halos into larger and larger halos. Some of these mergers 
involves sizeable clumps, most involves a more quiescent accretion of matter 
and small clumps from the surroundings. This process leaves its mark on the 
phase-space structure of the halos. Indeed, these dark halo streams are a major
source of attention in present day studies of the formation of our Galaxy 
\citep{helmi99,helmi00}. 

It remains an interesting question as to whether we can find evidence for these
merging events in the Fundamental Plane. \cite{cesar03a} look into the effects 
of major mergers on the Fundamental Plane and found that the Fundamental Plane 
does remain largely intact in the case of two merging ellipticals. However, 
what the effects will be of an incessant bombardment of a halo by material in 
its surroundings has not been studied in much detail. Given that this is a 
sensitive function of the cosmological scenario, we will study the influence on
FP parameters and thickness in more detail.

In this paper we address the specific question as to whether we can trace an 
influence of cosmic parameters in the scaling relations for simulated clusters,
and in particular the influence of the cosmic density parameter $\Omega_m$ and 
the cosmological constant $\Lambda$. We use a set of dissipationless $N$-body 
simulations involving open, flat and closed Universes. All the simulations are 
variants of the cold dark matter (CDM) scenario, representing different 
cosmologies, concerning both different values for the mass density 
$\Omega_{m}$, for dark energy $\Omega_{\Lambda}$ and for the implied power 
spectrum of density perturbations and the related merging and accretion history
of the clusters. 

The organization of this paper is as follows. In section \ref{sec5:sim} we 
describe the simulations and the definitions of the various parameters we use.
In section \ref{sec5:background}  we present a general description of the 
scaling relations which we investigate in this study before specifying the way 
in which we analyze them from the cluster-sized halos in our simulation. We 
investigate the scaling relations of galaxy clusters in different cosmologies 
at $z=0$ in section \ref{sec5:scaling}. Section ~\ref{sec5:evol_scaling} 
addresses the evolution of the scaling relations as a function of redshift and 
cosmic time. We also investigate the dependence of merging and accretion on the
scaling relations, which we discuss in section \ref{sec5:mergacc}. The 
interpretation of our results on the Fundamental Plane within the context of 
the virial theorem is discussed in section~\ref{sec5:reconcile}. Conclusions 
are presented in section \ref{sec5:conclusions}.

\section{The Simulations}
\label{sec5:sim}
We perform thirteen N-body simulations that follows the dynamics of $N=256^{3}$
particles in a periodic box of size $L=200h^{-1}$Mpc. The initial conditions 
are generated with identical phases for Fourier components of the Gaussian 
random field. In this way, each cosmological model contains the same 
morphological structures. For all models we chose the same Hubble parameter, 
$h=0.7$, and the same normalization of the power spectrum, $\sigma_{8}=0.8$.  
The principal differences between the simulations are the values of the matter 
density and vacuum energy density parameters, $\Omega_{m}$ and 
$\Omega_{\Lambda}$. By combining these parameters, we get models describing the
three possible geometries of the Universe: open, flat and closed. The effect of
having the same Hubble parameter and different cosmological constants 
translates into having different cosmic times. Table~\ref{table5:paramsim} 
lists the values of the cosmological parameters and the cosmic times at which 
the data is analysed.

The initial conditions are evolved up to the present time ($z=0$) using the 
massive parallel tree N-body code GADGET2 \citep{springel05}. The 
Plummer-equivalent softening was set at $\epsilon_{pl}=15h^{-1}$kpc in physical
units from $z=2$ to $z=0$, while it was taken to be fixed in comoving units at 
higher redshifts. For each cosmological model we wrote the output of 100 
snapshots, from $a_{exp}=0.2$ ($z=4$) to the present time, $a_{exp}=1$ ($z=0$), 
equally spaced in $\log(a)$.

\begin {table}
  \begin {center}
    \begin {tabular}{||c|c|c|c|c||}
      \hline
      \hline
      \hline
      Cosmic Time & SCDM & $\Lambda$CDMO2 & $\Lambda$CDMF2 & $\Lambda$CDMC2 \\
      \hline
      \hline
      2.36 & 1.49 & 4    & 2.71 & 2.21 \\
      3.26 & 1.01 & 2.92 & 1.98 & 1.60 \\
      4.06 & 0.74 & 2.35 & 1.56 & 1.24 \\
      5.07 & 0.50 & 1.83 & 1.19 & 0.93 \\
      9.31 & 0    & 0.71 & 0.38 & 0.24 \\
      \hline
      \hline
    \end {tabular}
    \caption{\small{Cosmic times in Gyr and the corresponding redshift for a 
set of four reference cosmological models.}}
    \label{table:ct}
  \end {center}
\end {table}

\begin{figure*}
\centering                                     
\includegraphics[width=0.21\textwidth]{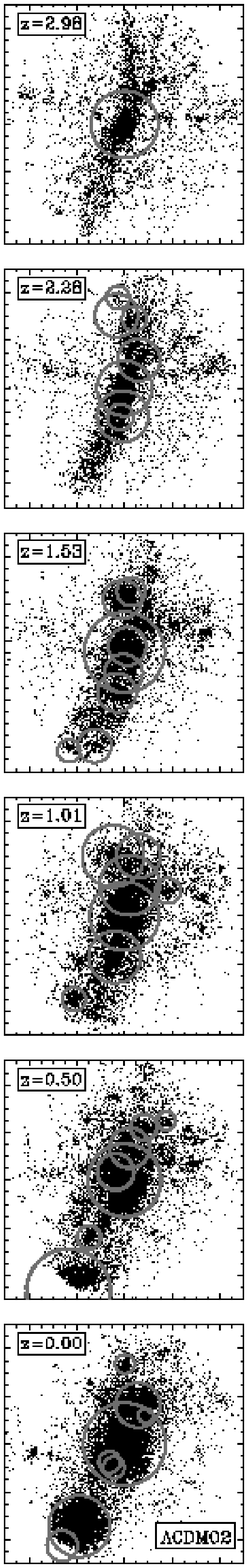}
\includegraphics[width=0.21\textwidth]{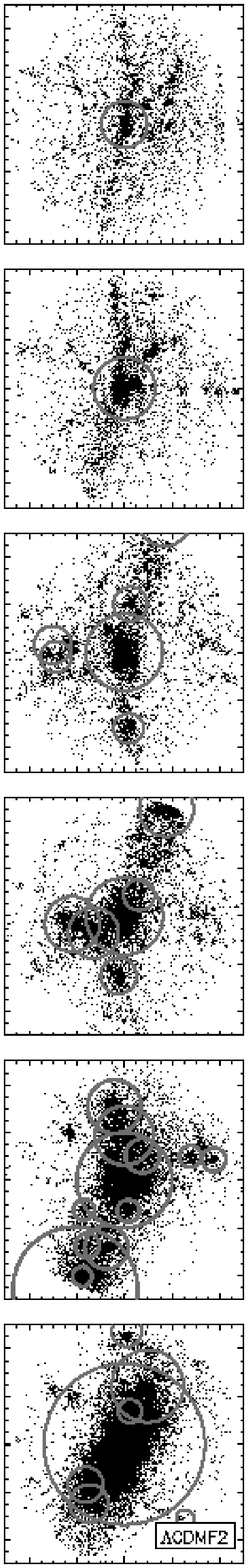}
\includegraphics[width=0.21\textwidth]{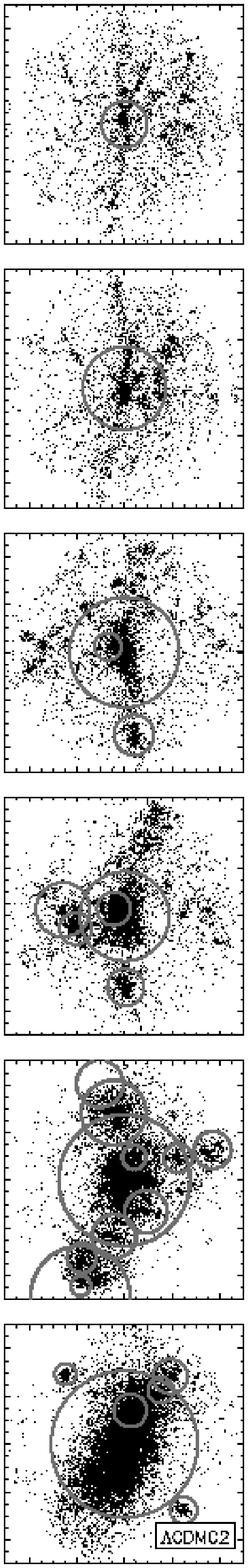}
\includegraphics[width=0.21\textwidth]{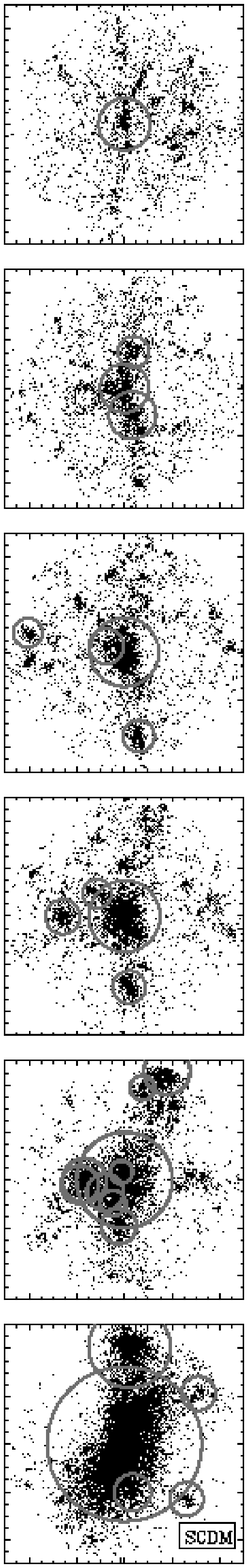}
\caption{{\small  Cluster evolution: $\Omega_{m}$ influence. Evolution as a function of redshift of a single dark matter halo in different cosmological models:  $\Lambda$CDMO2, $\Lambda$CDMF2, $\Lambda$CDMC2 and SCDM. Shown is the dark matter particle distribution in a box of comoving size $5h^{-1}$ Mpc, at 6 different redshifts: z=2.98, z=2.28, z=1.53, z=1.01, z=0.50 and z=0.00. The circles correspond to halos identified by HOP, with the size of the circle being proportional to their virial radius.}}
\label{fig:cluster_z}
\end{figure*}

\subsection{Halo identification}
\label{sec:haloid}
We use the HOP algorithm \citep{eisenstein98} to extract the groups present in 
the simulations. HOP associates a density to every particle. In a first step, a
group is defined as a collection of particles linked to a local density 
maximum. To make a distinction between a high density region and its 
surroundings, HOP uses a regrouping procedure. This procedure identifies a 
group as an individual object on the basis of a specific density value.  For 
this critical value we chose the virial density value $\Delta_{c}$ based on the
spherical collapse model.  In order to have the proper $\Delta_{c}$ we 
numerically compute its value for each of the cosmologies.  See 
Table~\ref{table5:paramsim} for the values of the virial density for each 
cosmology at $z=0$. For the latter we list two values: the virial overdensity 
$\Delta_{vir,b}$ with respect to the background density $\rho_{b}$ of the 
corresponding cosmology, and the related virial overdensity $\Delta_{vir,c}$ 
with respect to the critical density.

Note that we only consider groups containing more than 100 particles. Because  
the particle mass depends on the cosmological scenario, this implies a 
different mass cut for the halos in each of our simulations. As a result, SCDM 
does not have groups with masses lower than $10^{13}h^{-1}$M$_{\odot}$.  We have
to keep in mind this artificial constraint when considering collapse and 
virialization in hierarchical scenarios at high redshifts, and also when making
fits to the relationships among the various cluster parameters. In cases where 
structure growth is still continuing vigorously at the current epoch, the 
collapsed halos at high redshifts will have been small: our simulations would 
not be able to resolve these.

\subsection{Halos and Cosmology: an example}
Figure~\ref{fig:cluster_z} follows the evolution of one particular cluster halo
in four different cosmologies. These are $\Lambda$CDMO2, $\Lambda$CDMF2, 
$\Lambda$CDMC2 and SCDM. By using the same Fourier phases to set up the initial
conditions in each of the cosmologies we get a sample of corresponding halos. 
In each of the cosmologies the evolution of the cluster halo is shown at six 
different redshifts, from $z\approx 3$ onward to the present epoch $z=0$. The 
panels show the mass distribution in and around the cluster, and its 
progenitors, in a box of comoving size $5h^{-1}$ Mpc. Circles enclose halos 
identified by HOP, with the circle radius proportional to the virial radius of 
the group (i.e. the distance from the center of mass to the outermost particle 
of the group). Projection effects may occasionally cause circles to appear 
within circles. 

In all four cosmologies, the buildup of the halo clearly involves the merging 
of several smaller mass clumps, some of which are identified as genuine 
proto-halos by means of circles. Fig.~\ref{fig:cluster_z} shows that the 
sequence $\Lambda$CDMO2, $\Lambda$CDMF2, $\Lambda$CDMC2 and SCDM corresponds to
a sequence in which the formation of the halo shifts to later and later epochs.
At all depicted redshifts, and in particular at higher redshifts,  the clusters
in the $\Lambda$CDMO2 cosmology have considerably more pronounced and developed
mass concentration. 

\subsection{Halo properties}
\label{subsec:hp}
In our study, we limit ourselves to cluster-like halos. A galaxy cluster is 
defined as a dark matter halo with a mass M$>10^{14}h^{-1}$M$_{\odot}$. We 
measure three quantities for each cluster and test their scaling relations.

Scaling relations of collapsed and virialized objects relate two or three 
fundamental characteristics of those objects. The first involves a quantity 
measuring the amount of mass, often in terms of the amount of light emitted by 
the object. The second quantity involves the size of the object, while the 
third one quantifies its dynamical state. 

\begin{itemize}
\item \textbf{Mass:} defined as the number of particles multiplied by the mass 
per particle present in each group:
\begin{equation}
M = n_{part}m_{part}\,,
\label{eq:mmass}
\end{equation}
where $n_{part}$ is the number of particles in the halo and $m_{part}$ is the 
mass of each particle (see column $m_{dm}$ in Table~\ref{table5:paramsim}). The
mass of the particle is different for each cosmology. \\

\item \textbf{Surface mass density:} Alternatively, following observational 
practice, we use the magnitude-scale surface mass density $\mu$ for our 
Fundamental Plane evaluations, 
\begin{equation}
\mu\,=\,-2.5\log{M}\,+\,5\log{r}\,
\label{eq:mu}
\end{equation}
where $M$ and $r$ are the mass and the radius of the halo. Combining this with 
a mass to light ratio it becomes a surface brightness, one of the observables 
of the Fundamental Plane. \\

\item \textbf{Velocity dispersion:} computed as
\begin{equation}
\sigma_v^{2}=\frac{2K}{n_{part}m_{part}}\,,
\label{eq:sigmav}
\end{equation}
where $K$ is the kinetic energy of the halo.
\end{itemize}
\noindent As a measure for the size of the halos, we have explored two options:
the half-mass radius and the mean harmonic radius.

\begin{itemize}

\item \textbf{Half-mass radius:} $r_{half}$ is the radius that encloses half of
the mass of the clump. This radius is closest in definition to the half-light 
radius used in observational studies.\\

\item \textbf{Mean harmonic radius:} $r_{h}$ is defined as the inverse of the 
mean distance between all pairs of particles in the halo:
\begin{equation}
\frac{1}{r_{h}} = \frac{1}{N}\sum_{i<j}\frac{1}{|\mathbf{r}_{ij}|}\,,
\hspace{5mm}N=\frac{n_{part}(n_{npart}-1)}{2}\,,
\label{eq5:mrh}
\end{equation}
where $\mathbf{r}_{ij}$ is the separation vector between the $i$th and the 
$j$th particle. The great virtue of this radius is that it is a good measure of
the effective radius of the gravitational potential of the clump, certainly 
important when assessing the virial status of the clump. Also, it has the 
practical advantage of being independent of the definition of the cluster 
center. To some extent, it is also an indicator of the internal structure of 
the halo because it put extra weight to close pairs of particles.
\end{itemize}

\begin{figure}
\centering
\vskip -0.5truecm
\includegraphics[width=0.47\textwidth]{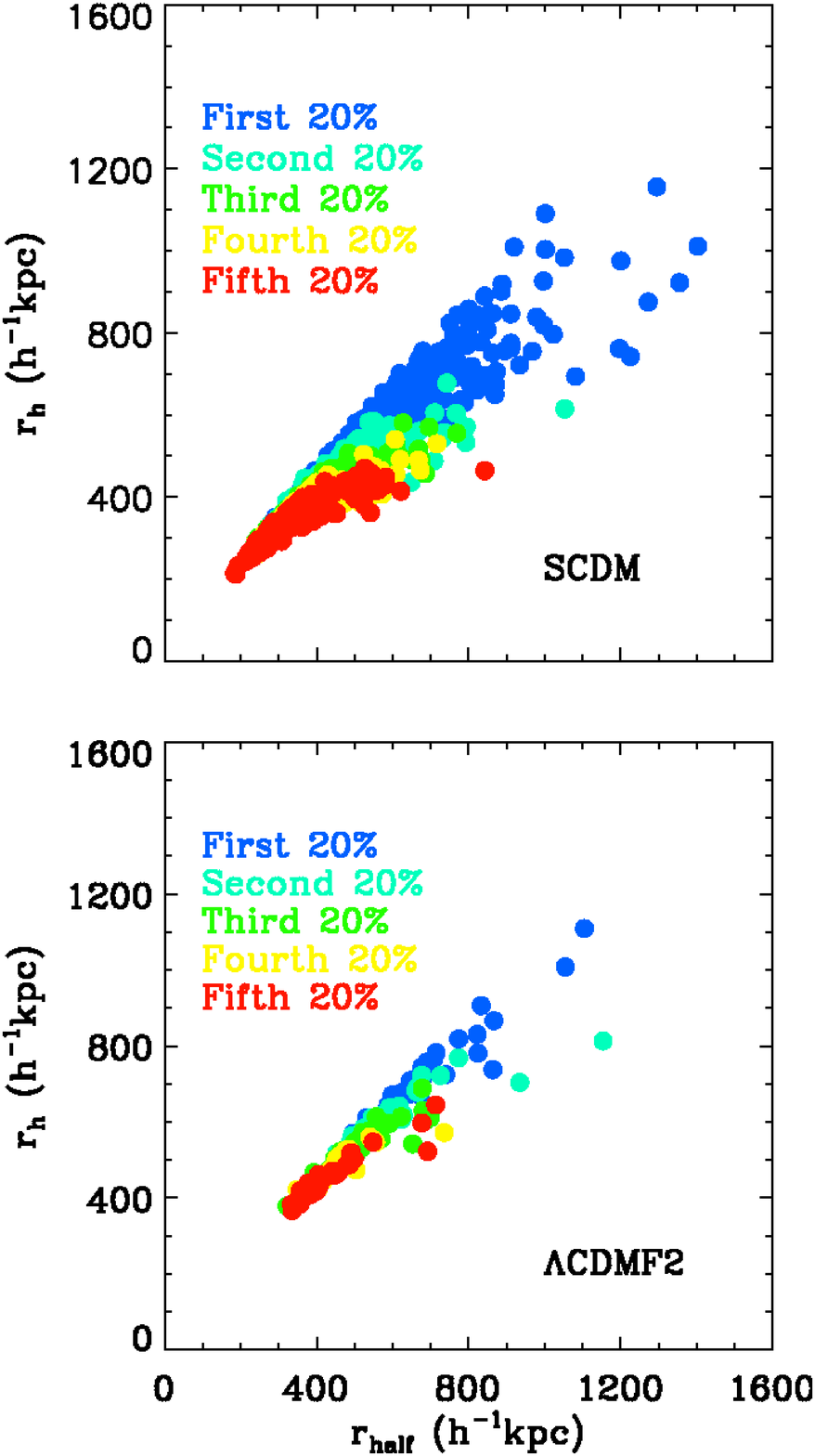}
\caption{\small{Comparison between the mean harmonic radius and the half-mass
radius of the cluster-size halos in the SCDM (top) and $\Lambda$CDMF2 (bottom) 
scenarios. The colours depict different mass ranges, each colour representing a
20 percentile mass quantile.}}
\label{fig:radiusrel}
\end{figure}

Most of the results presented in this paper refer to the mean harmonic radius 
of the halos: this seems rather natural given that we are discussing the virial
theorem (see Table \ref{table:scaling_harm}). We have also compared the results
obtained using the half-mass radii of the halos. 

In Fig. \ref{fig:radiusrel} we plot the mean harmonic radius versus the 
half-mass radius of the cluster-size halos in the SCDM and in the 
$\Lambda$CDMF2 models. We see that the relationship is not very tight at larger
masses, and that the differences between the two radii are particularly 
prominent in the SCDM cosmology. We shall discuss this further in 
section~\ref{sec5:mergacc}. Not surprisingly, the fitted Fundamental Plane 
parameters depends strongly on which radius is used.  Equally surprising, the 
Kormendy relation slope does not seem to be particularly sensitive to the 
choice of $r_h$ or $r_{half}$ (the slopes are statistically not different). 
This is summarised in Table \ref{sec5:mergacc}.  

\section{Scaling Relations}
\label{sec5:background}
For the samples of cluster-sized halos in our simulations we will be assessing 
three specific scaling relations: the Faber-Jackson relation, the Kormendy 
relation and the Fundamental Plane. 

From observations of elliptical galaxies we have learned that there are tight 
scaling relations between a few of their fundamental structural properties (see
e.g. \cite{binney98}). These properties are the total luminosity $L$ of a 
galaxy - or its surface brightness $\mu$ - its characteristic size $R_{e}$ and 
its velocity dispersion $\sigma_{v}$. 

\subsection{Faber-Jackson and Kormendy Relations}
The first scaling relation is the Faber-Jackson relation \citep{fj76} between 
the luminosity $L$ of the galaxy and its velocity dispersion $\sigma_v$, 
\begin{equation}
L\propto\sigma_{v}^{\beta}\,,
\label{eq:fjrel}
\end{equation}
where the index $\beta\sim4$. A similar relation, known as the Tully-Fisher 
relation \citep{tully77}, holds for HI disks of spiral galaxies. According to 
this relation, the galaxies' rotation velocity is tightly correlated with the 
absolute magnitude of the galaxy.

Another relation was established by \cite{kormendy77}. He found that there is a
strong, not entirely unexpected, correlation between the luminosity $L$ and 
effective radius $R_e$ of the elliptical galaxies: 
\begin{equation}
L\propto R_{e}^{\alpha}\,,
\label{eq:korrel}
\end{equation}
where the index $\alpha\sim1.5$. 

\begin{table*}
  \begin {center}
    \begin{tabular}{|l|l|l|c|c|c|c|c|c|c|c|c|c|c|c|c|c|r|}
      \hline
             &&            &                    &$r_{h}\propto M^{a}$ &&&
       $\sigma\propto M^{b}$ &&&\multicolumn{6}{c}{$\log r_{h}\,=\, c \mu+d \log \sigma$+$C_{fp}$}\\[0.6ex]
       Model & $\Omega_{m}$ & $\Omega_{\Lambda}$ & 
       & $a$ & $\mathcal{S}_K$ & 
       & $b$ & $\mathcal{S}_{FJ}$ &&
       & $c$ &$\sigma_{c}$ & $d$ & $\sigma_{d}$ & $\mathcal{S}$ & $w_{fp}$\\[0.6ex]
       & & &
       & && 
       && &&
       && ($10^{-2}$) & & ($10^{-2}$) & & ($10^{-3}$)\\[0.6ex]
      \hline
      \hline
       SCDM           & 1   & 0   && 0.38 & 0.06 && 0.33 & 0.03 &&& 
       0.37 & 0.31 & 1.78 & 1.14 & 0.03 & 14.03 \\
       OCDM01         & 0.1 & 0   && 0.34 & 0.05 && 0.37 & 0.03 &&&
       0.35 & 1.73 & 1.60 & 8.27 & 0.02 & 8.57\\
       OCDM03         & 0.3 & 0   && 0.36 & 0.05 && 0.33 & 0.03 &&&
       0.38 & 1.10 & 1.76 & 4.13 & 0.03 & 13.41\\
       OCDM05         & 0.5 & 0   && 0.37 & 0.05 && 0.33 & 0.03 &&&
       0.37 & 0.53 & 1.79 & 2.01 & 0.03 & 12.24\\
       $\Lambda$CDMO1 & 0.1 & 0.5 && 0.35 & 0.05 && 0.38 & 0.03 &&&
       0.37 & 1.55 & 1.60 & 7.56 & 0.02 & 7.51\\[0.5ex]       
       $\Lambda$CDMO2 & 0.1 & 0.7 && 0.38 & 0.05 && 0.35 & 0.03 &&&
       0.38 & 1.90 & 1.66 & 8.84 & 0.02 & 8.44\\[0.5ex]
       $\Lambda$CDMF1 & 0.1 & 0.9 && 0.35 & 0.05 && 0.36 & 0.03 &&&
       0.38 & 1.64 & 1.69 & 7.93 & 0.01 & 6.97\\[0.5ex]
       $\Lambda$CDMO3 & 0.3 & 0.5 && 0.36 & 0.05 && 0.34 & 0.03 &&&
       0.41 & 1.03 & 1.86 & 3.88 & 0.03 & 11.43\\[0.5ex]
       $\Lambda$CDMF2 & 0.3 & 0.7 && 0.36 & 0.05 && 0.33 & 0.03 &&&
       0.41 & 1.11 & 1.88 & 4.01 & 0.02 & 11.23\\[0.5ex]
       $\Lambda$CDMC1 & 0.3 & 0.9 && 0.34 & 0.05 && 0.34 & 0.03 &&&
       0.42 & 1.19 & 1.92 & 4.33 & 0.03 & 11.09\\[0.5ex]
       $\Lambda$CDMF3 & 0.5 & 0.5 && 0.35 & 0.05 && 0.34 & 0.03 &&&
       0.38 & 0.56 & 1.81 & 2.10 & 0.03 & 11.75\\[0.5ex]
       $\Lambda$CDMC2 & 0.5 & 0.7 && 0.35 & 0.05 && 0.34 & 0.03 &&&
       0.38 & 0.57 & 1.82 & 2.14 & 0.03 & 11.66\\[0.5ex]
       $\Lambda$CDMC3 & 0.5 & 0.9 && 0.34 & 0.05 && 0.34 & 0.03 &&&
       0.39 & 0.61 & 1.83 & 2.31 & 0.03 & 12.24\\[0.5ex]      
      \hline
      \hline
    \end{tabular}
    \caption{\small{Parameters of the scaling relations derived for the galaxy 
clusters in each of the simulated cosmological simulations.  $r_h$ is the mean 
harmonic radius of the cluster.    $a$ is the scaling parameter for the 
Kormendy relation, $b$  is the scaling parameter for the Faber-Jackson relation
and $c$ and $d$  are the scaling parameters for the Fundamental Plane. 
$\sigma$ is the standard error in each of the scaling relation parameters, 
$\mathcal{S}$ is the corresponding standard error/significance of the fit.}}
    \label{table:scaling_harm}
  \end {center}
\end {table*}

\subsection{Galaxy Fundamental Plane}
Both the FJ and Kormendy relations relate two structural characteristics and 
should be seen as projections of a more fundamental and tight relation between 
all three structural quantities: the \emph{Fundamental Plane} (FP). The 
Fundamental Plane of elliptical galaxies was first formulated by 
\cite{djorgovski87} and \cite{dressler87}. When we take the three-dimensional 
space defined by the effective radius $R_{e}$ of the galaxy, its surface 
brightness $I_{e}$ (with total luminosity $L\propto I_{e}R_{e}^{2}$) and 
velocity dispersion $\sigma_{v}$, we find that they do not fill space 
homogeneously but instead define a thin plane.

In logarithmic quantities, this plane may be parameterized as
\begin{equation}
 \log R_e\,=\,\gamma \log I_e\,+\, \delta \log \sigma_v\,+\,C_{fp}
\end{equation}
For example, \cite{jorgensen96} found that a reasonable fit to the Fundamental 
Plane is given by
\begin{equation}
\log R_e\,=\,-0.82 \log I_e\,+\,1.24 \log \sigma_v\,+\,C_{fp}
\label{eq:jorgensen}
\end{equation}

While nearly all galaxies, ranging from giant ellipticals to compact dwarf 
ellipticals, appear to lie on the FP \citep[also see e.g.][]{jorgensen95, 
bernardi03, cappellari06, bolton07} it is interesting to note that diffuse 
dwarf ellipticals do not \citep{kormendy87}: they seem to be fundamentally 
different objects.  

The observed Fundamental Plane not only provides information on the dynamical 
state of the object but also on the evolution of its stellar content and, by 
implication, about its formation. For a virialized object with effective radius
$R_e$ and mass-to-light ratio $M/L$ the FP relation will have the form 
\begin{equation}
\log{R_e}\,=\,-\,\log{I_e}\,+\,2\log{\sigma_{v}}\,-\,\log{(M/L)}\,+\,C_{s}\,,
\end{equation}
in which $I_e=L/4\pi R_e^{2}$ is the mean surface brightness and $C_{s}$ a 
constant dependent on the structure of the object. 

The observed parameter values for elliptical galaxy Fundamental Plane (see 
Eqn.~\ref{eq:jorgensen}) are different from what might be expected for a plane 
that results simply from  virialization and constant mass-to-light ratio.  One 
explanation for this difference is that galaxies may be structurally equivalent
while having a mass-dependent $M/L$ ratio.  That would imply a formation 
process involving a tight fine tuning of $M/L$.  Nevertheless, pursuing this 
view, the parameters inferred by \cite{jorgensen96} (Eqn.~\ref{eq:jorgensen}) 
would imply a mass-to-light ratio dependence on mass:
\begin{equation}
(M/L)\,\propto\,M^{0.25}\,,
\end{equation}
using $M\propto \sigma_v^2 R_e$ and $L \propto I_e R_e^2$ 
\citep[see e.g.][]{faber87}. Recent semi-analytical modelling of galaxy 
formation suggest a more complex relation between the mass-to-light ratio and 
luminosity, involving a minimum $M/L$ for galaxies with 
$M\approx10^{11}-10^{12}h^{-1}$M$_{\odot}$.    In the absence of any 
mass-to-light dependency, the discrepancy between the planes would have to be 
due to variations in the structure parameters of the galaxies.  

There is an intrinsic scatter of the FP that has been found for elliptical 
galaxies: this has not been completely explained and may be a manifestation of 
the formation process. 

A slightly different approach is used in the gravitational lensing study of 
\cite{bolton07}. These authors presented a new formulation of the FP using 
lensing data to replace surface brightness with surface mass density, arriving 
at the relationship of the form 
\begin{equation}
\log{R_e}\,=\,\gamma\log{I_e}\,+\,\delta\log{\sigma_{e2}}\,+\,C_{fp}\,,
\label{eq:fp_bolton_1}
\end{equation}
where $\sigma_{e2}$ is the velocity dispersion within half of the effective 
radius $R_{e}$, and
\begin{equation}
    \gamma = -0.78 \pm 0.13, \quad \delta = 1.50 \pm 0.32, \quad C_{fp} = 
    3.9 \pm 1.7.
\end{equation}
Furthermore they suggest that the scatter about the Fundamental Plane, derived 
from their data, correlates with their derived mass-to-light ratio for the 
galaxies in their sample.  The evidence is not strong though it is suggestive.

They also present an interesting alternative, which they refer to as the 
``Mass Plane'' (MP), in which they find the dependence of $\log(R_e)$ on 
$\log(\sigma_{e2})$ and surface mass density $\Sigma_{e2}$ within a radius 
$R_e/2$: 
\begin{equation}
\log{R_e}\,=\,\gamma_m \log \Sigma_{e2}\,+\,\delta_m \log \sigma_{e2}\,+
\,C_{fp,m}
\label{eq:fp_bolton_2}
\end{equation}
with
\begin{equation}
\gamma_m = -1.16 \pm 0.09, \,\delta_m = 1.77 \pm 0.14, \,C_{fp} = 7.8 \pm 1.0.
\end{equation}
Using surface mass density $\Sigma_{e2}$ within a radius $R_e/2$ in place of 
surface brightness $I_e$ removes one of the assumptions about the relationship 
between mass and light.

\subsection{Cluster Fundamental Plane}
If clusters were fully virialized objects with the same internal dynamics, they
 would necessarily lie on a universal Fundamental Plane in the 
Mass-velocity-radius space.  This was first addressed by \cite{schaeffer93}, 
who, using sample of 29 Abell clusters, discovered a FP relation in 
light-velocity-radius space : $L\propto R_{e}^{0.89}\sigma_{v}^{1.28}$.  This is 
equivalent to the relationship 
\begin{equation}
\log{R_{e}}\,=-\,0.90 \log{I_{e}}\,+\,1.15 \log{\sigma_{v}}\,+\,C_{fp},,
\label{eq:enacs}
\end{equation}
in which $I_{e}$ is a measure of the mean surface brightness of the cluster.  
The corresponding FJ relation is $L\propto R_{e}^{1.87}$ and the Kormendy 
relation is $L\propto R_{e}^{1.34}$. Similar numbers were inferred by  
\cite{lanzoni04}, $L\propto R_{e}^{0.90}\sigma_{v}^{1.31}$.  

In a project designed to test this further, \cite{adami98} found a FP 
relation for a sample of ENACS Clusters, though their fitted parameters 
were markedly different: $L\propto R^{1.19 \pm 0.14}\sigma^{0.91 \pm 
0.16}$. This is equivalent to the relationship
\begin{equation}
\log{R_{e}}\,=-\,(1.23 \pm 0.20)\log{I_{e}}+(1.12 \pm 0.11) 
\log{\sigma_{v}}\,+\,C_{fp}\,,
\label{eq:enacs}
\end{equation}
in which $I_{e}$ is the mean surface brightness of the cluster.Note that 
there are considerable systematic uncertainties in these values which are not 
reflected in the quoted error bars: these arise out of the profile fitting to 
the cluster.  The above fit to the data is based on fitting a King profile, 
(this gave the best fit to the data). 

In studies of simulated dark matter dominated galaxy clusters, we can study 
scaling relations that are similar to those inferred from observable 
quantities.  To infer these relations we base ourselves on the mass $M$ of the 
object. If the  selected objects have the same average density, we would expect
an equivalent Kormendy relation given by
\begin{equation}
M\propto R_{e}^{3}\,.
\label{eq:kormendy-like}
\end{equation}
Any difference in slope should be ascribed to a dependence of mean density 
$\langle\rho(R_{e})\rangle$ on the size $R_{e}$ of the object. The equivalent 
Fundamental Plane relation will be that of Eqn.~\ref{eq:fp_virial}, while the 
Faber-Jackson relation would then be
\begin{equation}
M\propto\sigma_{v}^{3}\,.
\label{eq:fj-like}
\end{equation}
Note that this is based on the assumption of constant mean density $\rho$ of 
the selected objects, in line with HOP overdensity criterion (see 
sec.~\ref{sec:haloid}).

\cite{lanzoni04} analyzed the N-Body cluster scaling relations on the basis of 
a sample of 13 massive dark matter halos identified in a high resolution 
$\Lambda$CDM $N$-body simulations.  They were able to confirm the existence of 
FP relations for the simulation dark matter clusters and also found that these 
have a slope that was significantly different from the galaxy FP slope, 
\begin{equation}
\log{R_{e}}\,=\,(0.44 \pm 0.02)\,\mu\,+\,(1.92 \pm 0.12) \log{\sigma_{v}}\,
+\,C_{fp}\,,
\end{equation}
with $\mu$ the surface mass density (Eqn.~\ref{eq:mu}). The difference in FP 
parameters between the dark matter halos and those inferred for the observed 
cluster sample (see above, Eqn.~\ref{eq:enacs}) formed a key aspect of their 
study. They suggest a mass dependent cluster 
$M/L$ ratio 
\begin{equation}
(M/L)\,\propto\,M^{0.8}\,.
\end{equation}
would be able to explain the observed cluster Fundamental Plane. Interestingly,
this is markedly different from that inferred for early type galaxies.  Of 
course there is no obvious reason why the Fundamental Plane for galaxies should
have any bearing on the Fundamental Plane for clusters.  Indeed, as we shall 
see for the ENACS sample, its FP parameters values seem to be irreconcilable 
with the virial theorem. 

\subsection{Determination of Scaling Relations}
For the sample of $N$ cluster-sized halos in each simulation we study the 
scaling relations between their size $r$, mass $M$ - or equivalent surface mass
density $\mu$ - and velocity dispersion $\sigma$ (note that $N$ is in general 
different for each cosmology). Given the inferred mass $M$ 
(Eqn.~\ref{eq:mmass}), velocity dispersion $\sigma_v$ (Eqn.~\ref{eq:sigmav}) 
and the mean harmonic radius $r_{h}$ (Eqn. \ref{eq5:mrh}) of the cluster halos,
we find the scaling relation parameters by linear fitting of the relations. 

Sample selection effects play a complex role in the analysis of real data 
samples (\cite{barbera03}).  Fortunately the issue is far simpler when 
analysing clusters found in N-Body models where the only selection criterion is
a mass cut-off imposed by the cluster finding algorithm.  We deal with that 
simply by making the Mass of the cluster the independent variable in all fits 
where relevant: this eliminates biases introduced through this object selection.

\begin{figure*}
\centering
\includegraphics[width=0.80\textwidth]{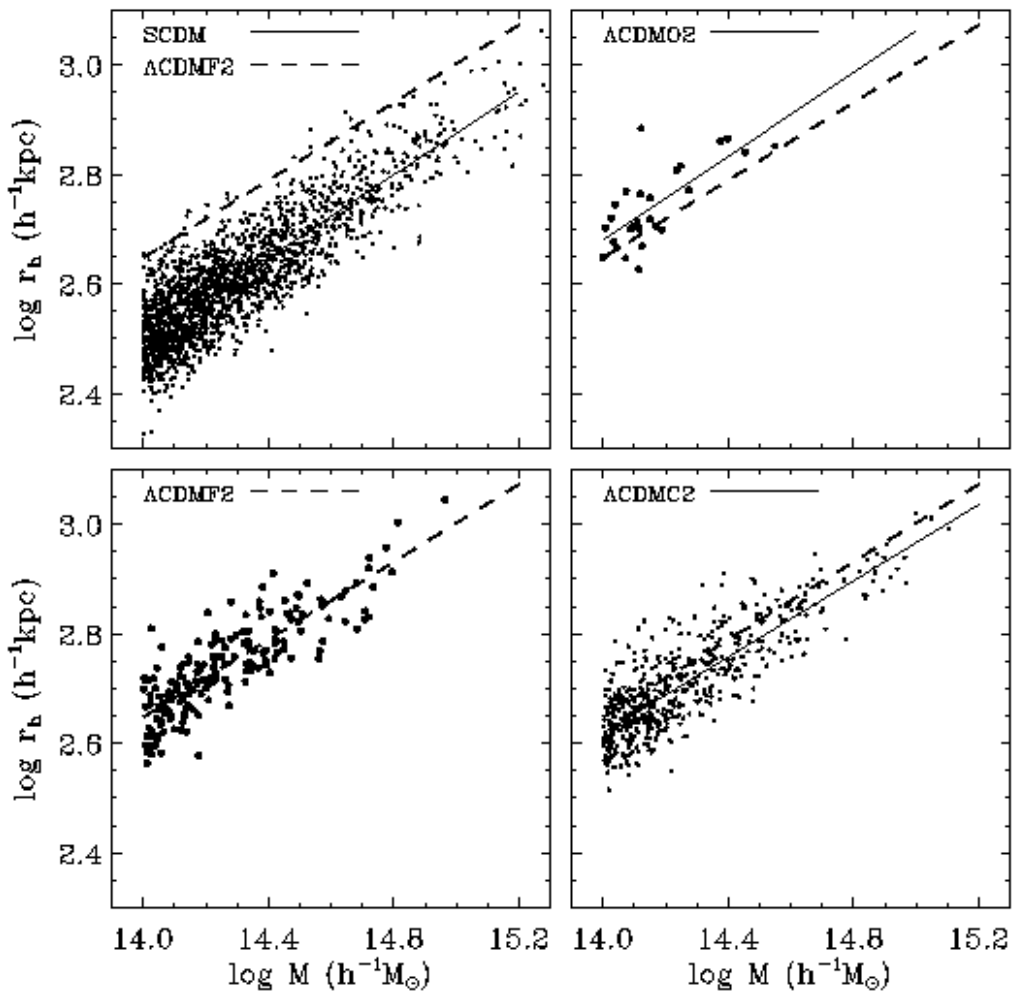}
\caption{\small{Kormendy Relation. Each panel plots the relation between mean 
harmonic radius $r_{h}$ and mass $M$ of the cluster-sized dark halos in the
simulations corresponding to one particular cosmology. Going from top left to 
bottom right these are: SCDM, $\Lambda$CDMO2, $\Lambda$CDMF2 and 
$\Lambda$CDMC2. In each of the panels we have superimposed the fitted Kormendy 
relation for the corresponding model and for $\Lambda$CDMF2 as comparison.}}
\label{fig:sr_korm}
\end{figure*}

\subsubsection{Kormendy Relation}
\noindent For the Kormendy relation we fit  
\begin{equation}
\log{r}\,=\,a\,\log{M}\,+\,C_{a}\,.
\label{eq:kormendy}
\end{equation}
to the $N$ data points $(\log{r_i},\log{M_i})$ of the halo sample. The 
significance $\mathcal{S}_K$ is computed from the $N$ residuals:
\begin{equation}
\mathcal{S}_K \,=\,\sqrt{\frac{1}{(N-1)}\,\sum_i^N\,(\log{r_i}-a\log{M_i}-C_a)2}
\label{eq:sk} 
\end{equation}

\subsubsection{Faber-Jackson Relation}
\noindent Along the same line, the Faber-Jackson relation is determined on the 
basis of the fit
\begin{equation}
\log{\sigma_v}\,=\,b\,\log{M}\,+\,C_{b}\,, 
\label{eq:fj}
\end{equation}
whose significance $\mathcal{S}_{FJ}$is calculated as follows:
\begin{equation}
\mathcal{S}_{FJ}\,=\,\sqrt{\frac{1}{(N-1)}\,\sum_i^N\,(\log{\sigma_{v,i}}-
b\log{M_i}-C_b)2} 
\label{eq:sfj}
\end{equation}

\begin{figure}
\centering
\includegraphics[width=0.41\textwidth]{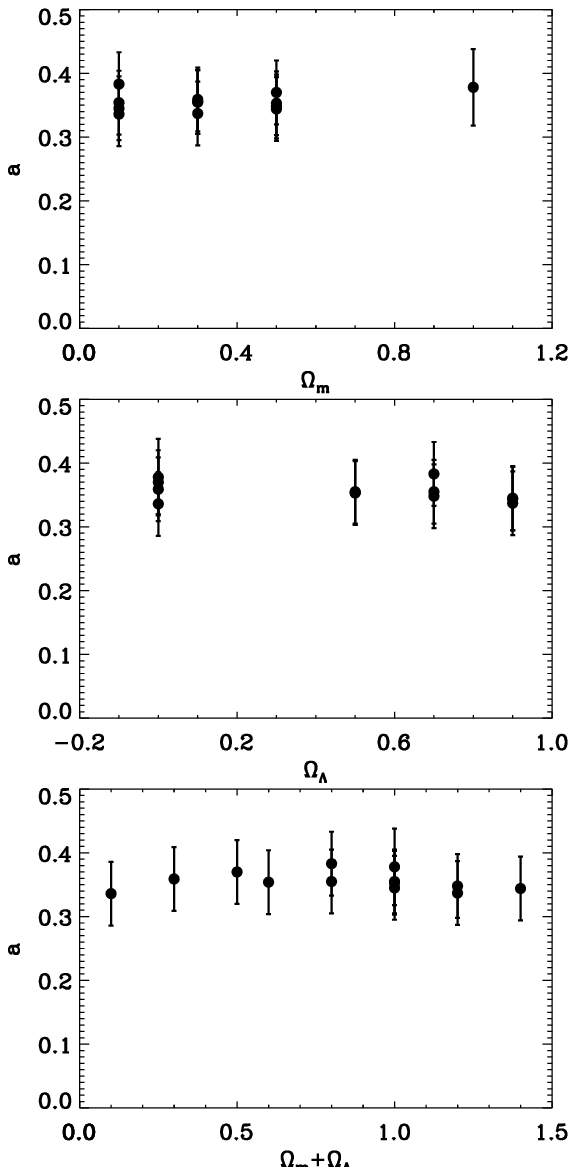}
\caption{{\small Inferred parameter $a$ for the Kormendy relation 
(Eqn.~\ref{eq:kormendy}) 
as a function of $\Omega_{m}$ (top panel), $\Omega_{\Lambda}$ (central panel) 
and $\Omega_{m}+\Omega_{\Lambda}$ (bottom panel). The bars represent the 
$1\sigma$ uncertainty range around the estimated parameter.}}
\label{fig:oms_mr}
\end{figure}
\begin{figure}
\centering
\includegraphics[width=0.41\textwidth]{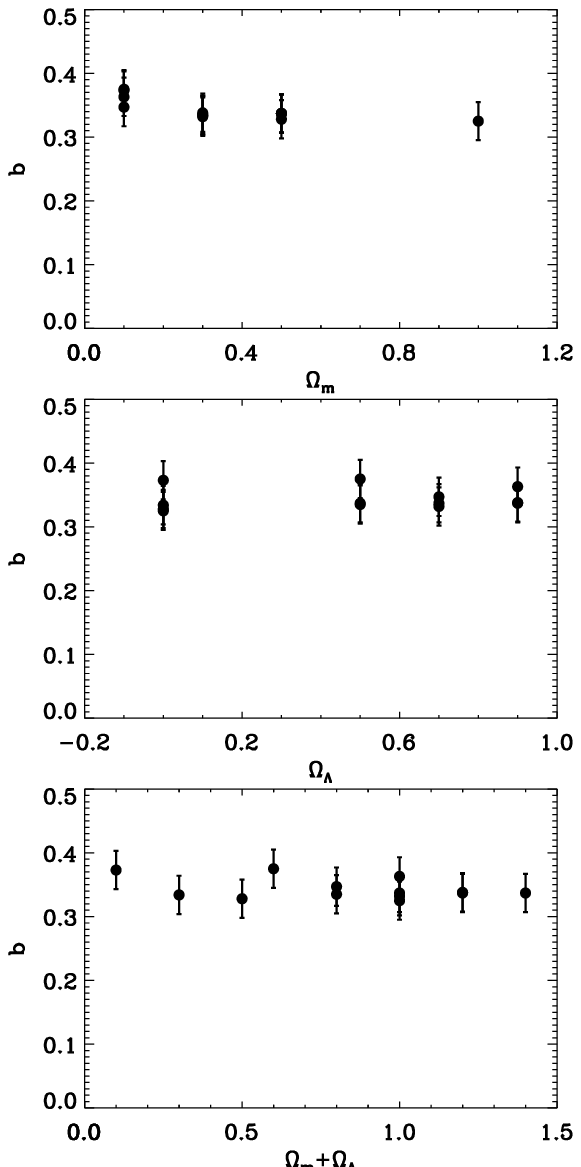}
\caption{{\small Inferred scaling parameter $b$ for the FJ relation 
as a function of three different parameters: $\Omega_{m}$ (top panel), 
$\Omega_{\Lambda}$ (central panel) and $\Omega_{m}+\Omega_{\Lambda}$ 
(bottom panel). The bars represent the $1\sigma$ uncertainty range around 
the estimated parameter.}}
\label{fig:oms_mv}
\end{figure}

\subsubsection{Fundamental Plane}
\noindent Instead of fitting the Fundamental Plane in the form of 
Eqn.~\ref{eq:fp_virial}, we do it in the way suggested by the observational 
work, i.e., using the surface mass density $\mu$ and velocity dispersion 
$\sigma_v$ as free parameters from which we determine a model for the radius, 
\begin{equation}
\log{r}\,=\,c\,\mu\,+\,d\,\log{\sigma_v}\,+C_{fp}\,.
\label{eq:fp}
\end{equation}
In this, $\mu$ is the magnitude-scale surface mass density (Eqn.~\ref{eq:mu}).
Although there are errors in determining both $\sigma_v$ and $\mu$, they are 
very small when compared with the dispersion about the Fundamental Plane. By 
fitting the parameters $c$ and $d$ this way we solve problems regarding biases 
in the mass (luminosity) selection. 

The significance $\mathcal{S}_{fp}$ of the Fundamental Plane fits derived from 
the sample of $N$ simulation cluster halos is computed according to:
\begin{equation}
\mathcal{S}_{fp} \,=\,\sqrt{\frac{1}{(N-2)}\,\sum_i^N\,(\log{r_i}-c\mu_i-\log{\sigma_v}-C_{fp})2}\,. 
\label{eq:sfp}
\end{equation}
The thickness $w_{fp}$ of the Fundamental Plane is estimated on the basis of 
the perpendicular distances of the cluster halos to the fitted plane:
\begin{equation}
\rm{w_{fp}}=\sqrt{{\frac{\sum{D_{\perp}^{2}}}{N}}}\,,
\end{equation}
where $N$ is the number of cluster halos in the sample and $D_{\perp}$ is the 
perpendicular distance of a point to a plane
\begin{equation}
D_{\perp} = \frac{c\mu+d\log{\sigma_v}+C_{fp}-
    {\log{r_h}}}{(c^{2}+d^{2}+1)^{1/2}}\,.
\end{equation}

\section{Scaling Relations in Different Cosmologies: \lowercase{z}=0}
\label{sec5:scaling}
We first investigate the scaling relations of the cluster dark matter halos in 
our cosmological models at the current epoch, $z=0$, and look for possible 
systematic differences between the parameter values and FP thickness as a 
function of the cosmology.  The parameters of the resulting linear fits, to be 
discussed in the following subsections, are listed in 
Table~\ref{table:scaling_harm}.

\subsection{Kormendy Relation}
Fig. \ref{fig:sr_korm} shows the relation between the mean harmonic radius 
$r_{h}$ of each cluster halo and their mass $M$. Each of the four panels 
depicts the relation for the halos in one particular simulated cosmology. The 
top left panel shows the SCDM cosmology, the top right one the $\Lambda$CDMO2 
model, the bottom left one the $\Lambda$CDMF2 model and the bottom right one 
the $\Lambda$CDMC2 model.

In each cosmology there is a strong and systematic almost linear relation 
between $\log{M}$ and $\log{r_{h}}$: the Kormendy relation appears to be a good
description for all situations.  A visual comparison between SCDM relation (top
left panel), the $\Lambda$CDMO2 relation (top right panel) and the 
$\Lambda$CDMF2 relation (bottom left panel) shows that clusters of comparable 
mass have a larger size in the low $\Omega_{m}$ cosmology than in the ones with
a higher density value. In other words, clusters are more compact in the SCDM 
cosmology.  Not unexpectedly we find objects of a higher density in higher 
$\Omega_{m}$ models. 

When fitting the plotted point distributions, we infer the parameter values 
listed in Table \ref{table:scaling_harm}. In each of the panels in 
Fig.~\ref{fig:sr_korm} we plotted the linear fits for all of the four depicted 
cosmologies. We find similar slopes for all cosmologies, in the order of 
$a\sim0.36-0.38$. This seems to imply that the mean density 
$\langle\rho(r_{h})\rangle \propto M^{-0.1}$: more massive halos have a 
slightly lower average density (see  also \cite{lanzoni04}). To investigate the
dependence of the Kormendy parameter $a$ on the cosmology in 
Fig.~\ref{fig:oms_mr} we have plotted the slope $a$ as a function of the 
average mass density parameter $\Omega_{m}$ (top panel), as a function of the 
cosmological constant $\Omega_{\Lambda}$ (central panel) and as a function of 
the cosmic curvature, in terms of $\Omega_{total}=\Omega_{m}+\Omega_{\Lambda}$  
(lower panel). There is no evidence for any systematic trends of the Kormendy 
parameter as a function of cosmology.  No evidence for an influence of either 
cosmic density $\Omega_m$ and $\Omega_{\Lambda}$ on the internal structure of 
the halos could be detected. 

\begin{figure*}
\centering
\includegraphics[width=0.80\textwidth]{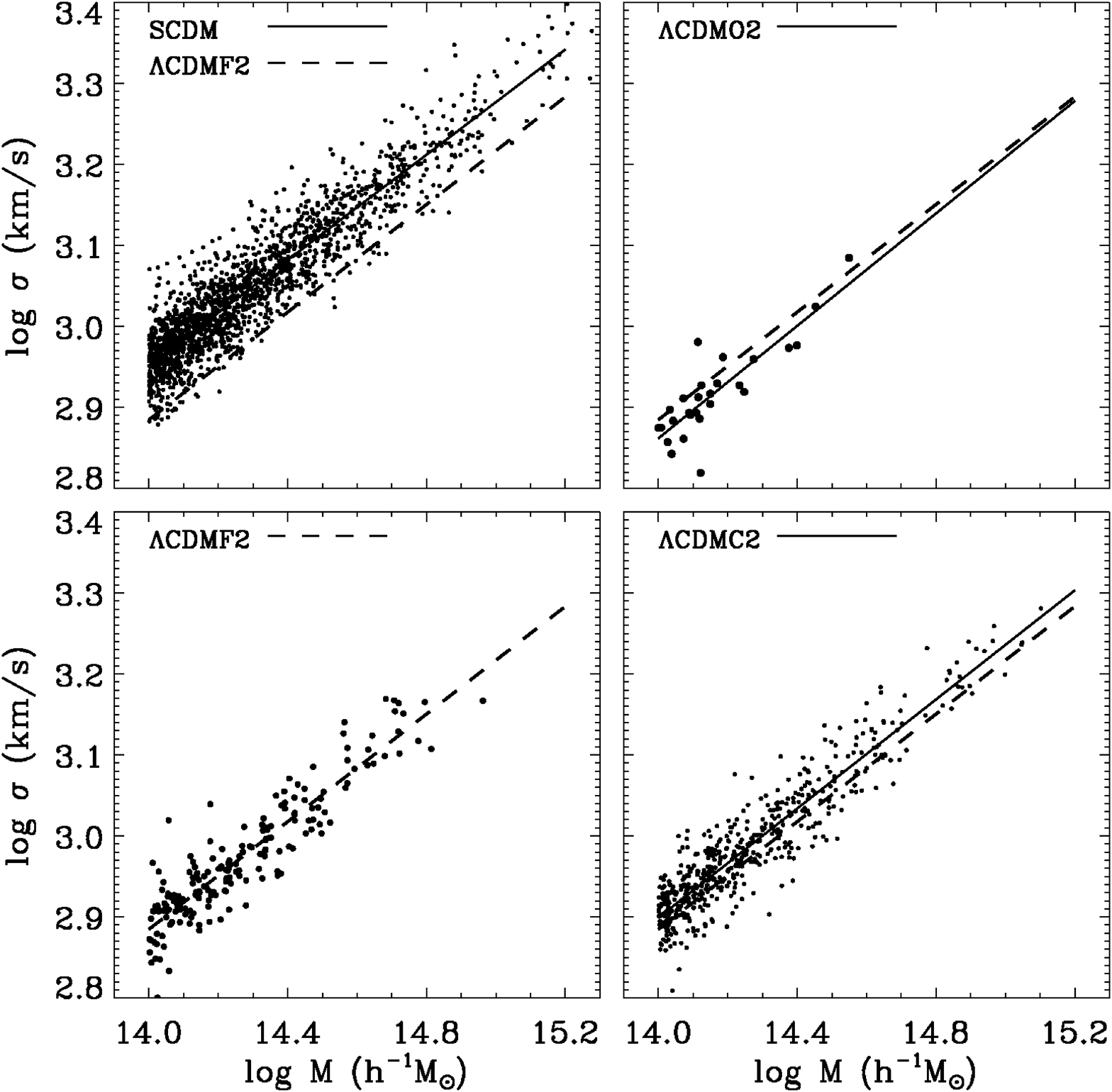}
\vspace{0.0cm}
\caption{\small{Faber-Jackson relation. Each panel plots the relation between 
the velocity dispersion $\sigma_v$ and the mass $M$ of the cluster-sized dark 
halos in the simulations corresponding to one particular cosmology. Going from 
top left to bottom right these are: SCDM, $\Lambda$CDMO2, $\Lambda$CDMF2 and 
$\Lambda$CDMC2. In each of the panels we have superimposed the fitted 
Faber-Jackson relation for the corresponding model and for $\Lambda$CDMF2 for
comparison.}}
\label{fig:sr_fj}
\end{figure*}
\subsection{Faber-Jackson Relation}
Fig.~\ref{fig:sr_fj} shows the Faber-Jackson relation: the relation between the
mass $M$ and the velocity dispersion $\sigma_v$ of the cluster halos.  Like in 
Fig. \ref{fig:sr_korm}, each of the four panels corresponds to one particular 
simulated cosmology: SCDM (top left panel), $\Lambda$CDMO2 (top right panel), 
$\Lambda$CDMF2 (bottom left panel) and $\Lambda$CDMC2 (bottom right panel). 

For comparison, in each of the panels we show the line of the $\Lambda$CDMF2 
model corresponding to the linear fit of this relation in each of the depicted 
cosmologies. The $M-\sigma_v$ relation is clearly well fitted by the 
Faber-Jackson like relation. It is considerably tighter than the equivalent 
Kormendy relation. 

It is also interesting is to note that, as with the Kormendy relation, we do 
not find any significant dependence of the FJ relation on the underlying 
cosmology: the slope $b$ in all cases is in the order of $b\sim 0.35$ (see 
Table~\ref{table:scaling_harm}). 
We also did not find any dependence on $\Omega_{\Lambda}$ or $\Omega_{total}$ 
(see Fig.~\ref{fig:oms_mv}). 

\begin{figure*}
\centering
\includegraphics[width=0.80\textwidth]{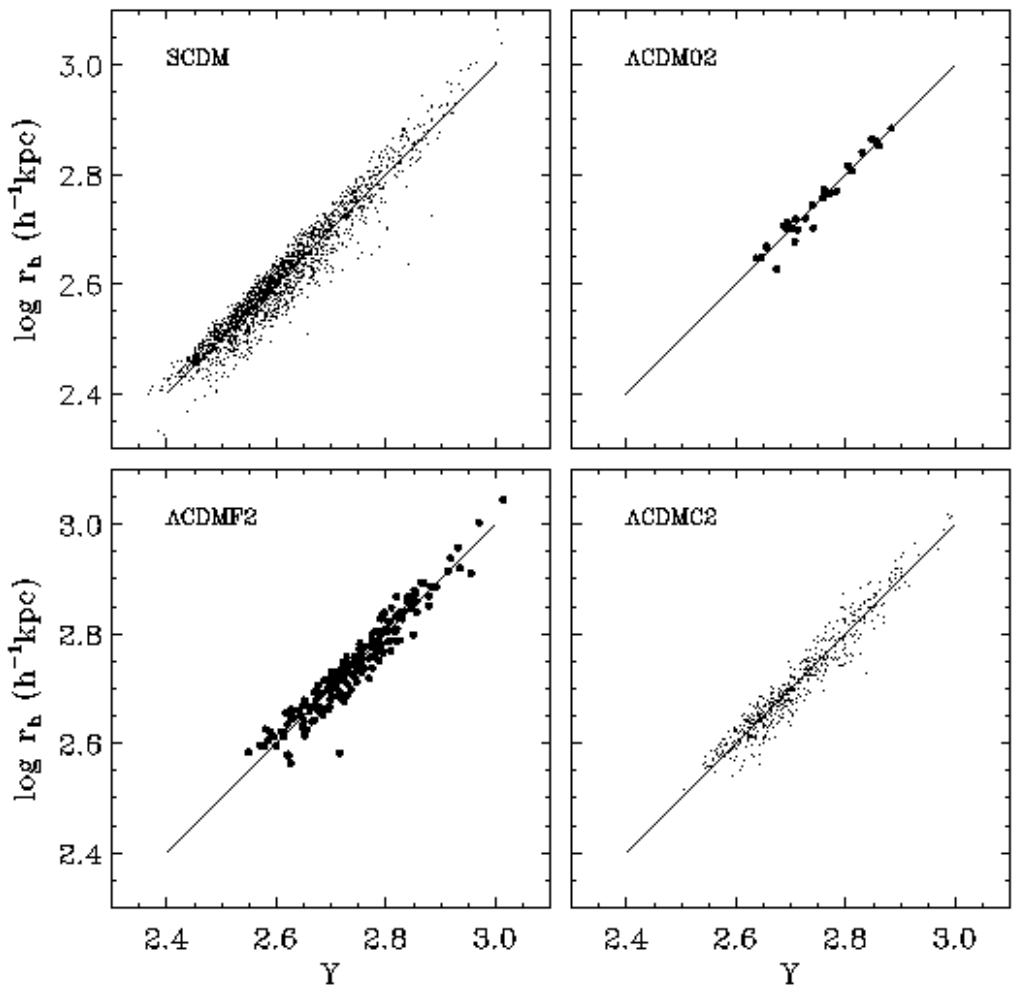}
\caption{\small{Fundamental Plane. Each panel plots the relation between 
harmonic radius $r_{h}$ and the quantity $Y=c\mu + d \log\sigma$. Combining 
the surface mass density $\mu$ and velocity dispersion $\sigma_v$, the scaling 
parameters $c$ and $d$ are the ones inferred from the FP fitting procedure. 
Top left panel: the relation between harmonic halo radius $r_h$ and $Y$ for 
the cluster halo sample in the SCDM simulation. Top right panel: for the 
halos in the $\Lambda$CDMO2 simulation. Bottom left panel: for the halos in 
the $\Lambda$CDMF2 simulation. Bottom right panel: for the halos in the 
$\Lambda$CDMC2. The superimposed lines in each panel represent the relation 
for the fitted Fundamental Plane for the corresponding cosmology. Note that, 
by definition, each of these fitted lines should have slope unity.}}
\label{fig:sr_fp}
\end{figure*}

Although the difference between the inferred value of $b\sim0.35$ in most 
cosmologies and the value of $b=0.33$ expected for virialized perfectly 
homologous systems (see Eqn.~\ref{eq:fj-like}) is not really significant, the 
consistent and systematic value $b>0.33$ might be suggestive for a weakly 
homologous population along the lines described in e.g. \cite{bertin02}.

\subsection{Fundamental Plane}
The Kormendy relation and the Faber-Jackson relation are two dimensional 
projections of an intrinsically three dimensional relation between mass $M$, 
size $r$ and velocity dispersion $\sigma_v$ of the halos. By implication, the 
spread of the Fundamental Plane relation should be less than that of each of
the two previous relations. 

The Fundamental Plane obtained for the same cosmologies as shown in 
Fig.~\ref{fig:sr_korm} and \ref{fig:sr_fj} (SCDM, $\Lambda$CDMO2, 
$\Lambda$CDMF2 and  $\Lambda$CDMC2) is illustrated in Fig.~\ref{fig:sr_fp}. 
In each of the frames we have plotted the harmonic radius $r_{h}$ of the halos 
against the quantity $Y=c  \mu + d \log \sigma_{v}+C_{FP}$ on a log-log plot. 
The parameters $c$ and $d$ in the latter quantity, $Y$, combining the surface 
mass density $\mu$ and the velocity dispersion $\sigma_v$ of each halo, are the
best fit FP parameters for the corresponding cosmology (see 
Table~\ref{table:scaling_harm}). 

The galaxy clusters in each cosmology do indeed seem to populate a tightly 
defined plane. The point clouds in each of the frames confirm our expectation 
that they should have a much lower scatter around the plane than in the case of
the Kormendy and Faber-Jackson relation (see Table~\ref{table:scaling_harm}).

From Table \ref{table:scaling_harm} we find a surprising level of consistency 
between the Fundamental Planes in each of the cosmologies. We find that the 
inferred parameters are close to the one theoretically expected for perfectly 
homologous virialized clusters halos. The inferred scaling parameter $c$ for 
the surface density $\mu$ hovers around $0.38-0.42$, close to the theoretical 
value $c \approx 0.4$ ($M\propto r_{h}\sigma_v^{2}$). The difference is 
somewhat larger for the parameter $d$, implying that the velocity dispersion 
scaling has a difference of $\sim0.15-0.25$ from the theoretical value of $2$.

\begin{figure*}
\centering
\vskip -0.2truecm
\includegraphics[width=0.80\textwidth]{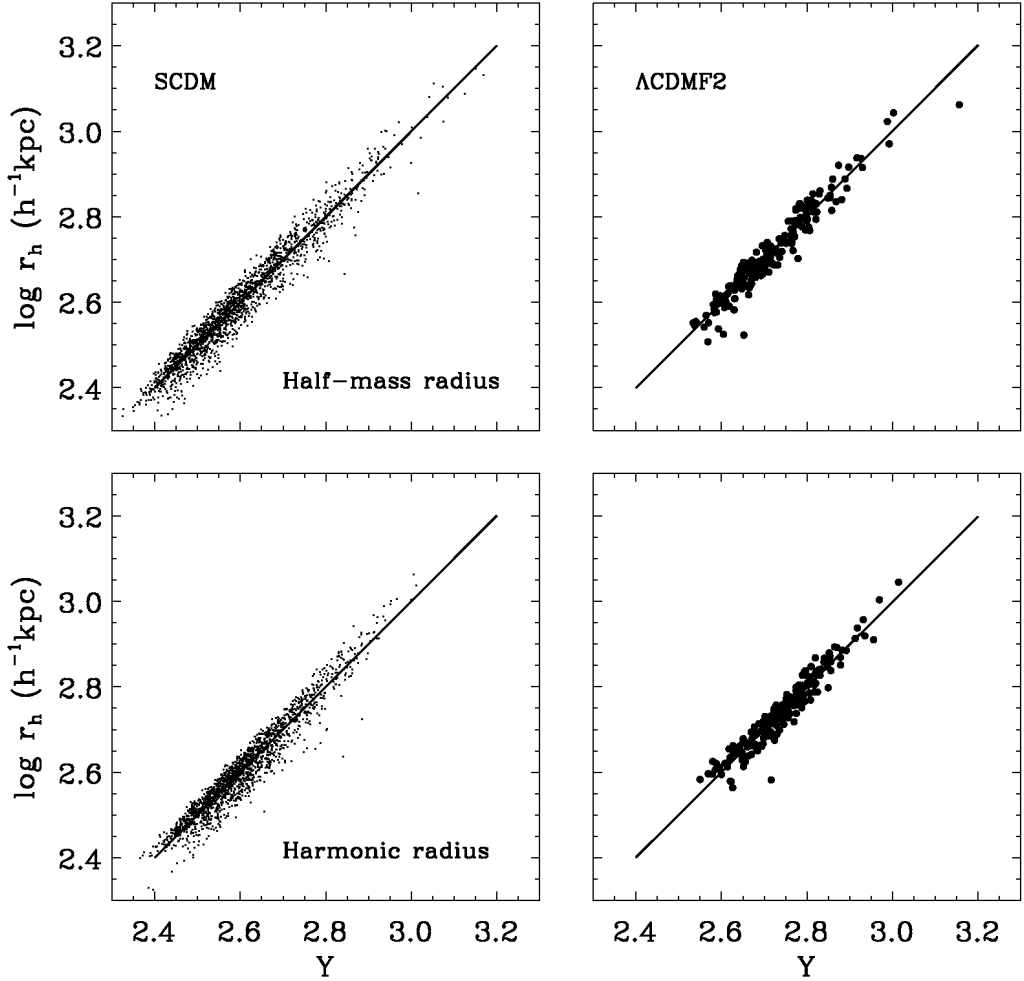}
\vskip -0.25truecm
\caption{\small{Fundamental Plane relation of dark halos in the SCDM (left) and 
$\Lambda$CDMO2 (right) cosmology, using the half-mass radius $r_{half}$ (top 
row) and the harmonic radius $r_{h}$ (bottom row). Plotted are mass $M$ versus 
the FP quantity $Y= c\mu + d \log\sigma$, with $c$ and $d$ the scaling 
parameters inferred from the (linear) fitting procedure. The lines represents 
the best fit FP relations.}}
\label{fig:fp_virhalf}
\end{figure*}

\begin{table*}
\vskip 0.2truecm
  \begin {center}
    \begin{tabular}{||l|l|l|l|l|c|c|c|c|c|c|c|c||}
      \hline
      \hline
            &              &                   &         &
      &\multicolumn{2}{c}{$r\propto M^{a}$} &&&\multicolumn{3}{c}{$\log r_{h} = c \mu+d \log \sigma_v$+$C_{fp}$} \\[0.6ex]
      Model & $\Omega_{m}$ & $\Omega_{\Lambda}$ && Radius &
      &$a$&$\mathcal{S}_K$&&$c$&$d$&$\mathcal{S}_{fp}$\\[0.6ex]
      \hline
      \hline
      &   &   && Half-mass &&0.39&0.09 &&0.29&1.60&0.03\\
      \raisebox{1.0ex}{SCDM} &\raisebox{1.0ex}{1} &\raisebox{1.0ex}{0} 
      && Harmonic &&0.38&0.06 &&0.37&1.78&0.03\\[0.6ex]
      \hline
      &  &  && Half-mass && 0.36 & 0.08&& 0.30 & 1.53 &0.02\\
      \raisebox{1.0ex}{$\Lambda$CDMO2}&\raisebox{1.0ex}{0.1}&
      \raisebox{1.0ex}{0.7}&&Harmonic&&0.38 & 0.05&&0.38&1.66&0.02\\[0.6ex]
      \hline
      &  &  && Half-mass && 0.35 & 0.07 && 0.31 & 1.66 & 0.03\\
      \raisebox{1.0ex}{$\Lambda$CDMF2} &\raisebox{1.0ex}{0.3}&
      \raisebox{1.0ex}{0.7}&& Harmonic&&0.36 & 0.05 && 0.41 & 1.88 & 0.02\\[0.6ex]
      \hline
      &  &  && Half-mass&& 0.35 & 0.08 && 0.30 & 1.63 & 0.03\\
      \raisebox{1.0ex}{$\Lambda$CDMC3} &\raisebox{1.0ex}{0.5}&
      \raisebox{1.0ex}{0.9} &&Harmonic&&0.35 & 0.05 && 0.38 & 1.82 & 0.03\\[0.6ex]
      \hline
      \hline
    \end{tabular}
    \label{table:halfvir}
  \end {center}
    \caption{\small{Scaling relation parameters and radius definition: inferred 
Kormendy relation parameter $a$ and Fundamental Plane parameters $c$ and $d$, 
based on the use of half mass radius $r_{half}$ and harmonic radius  
$r_{h}$. For four different cosmologies -- SCDM, $\Lambda$CDMO2, $\Lambda$CDMF2 
and $\Lambda$CDMC3 -- the scaling parameters and the corresponding 
goodness-of-fit ${\mathcal S}$ (see Eqn.~\ref{eq:sk} and Eqn.~\ref{eq:sfp}) are
given for $r_{half}$ (top row) and $r_{h}$ (bottom row).}}
\end {table*}

\begin{figure}
\centering
\mbox{\hskip -0.25truecm\includegraphics[width=0.41\textwidth]{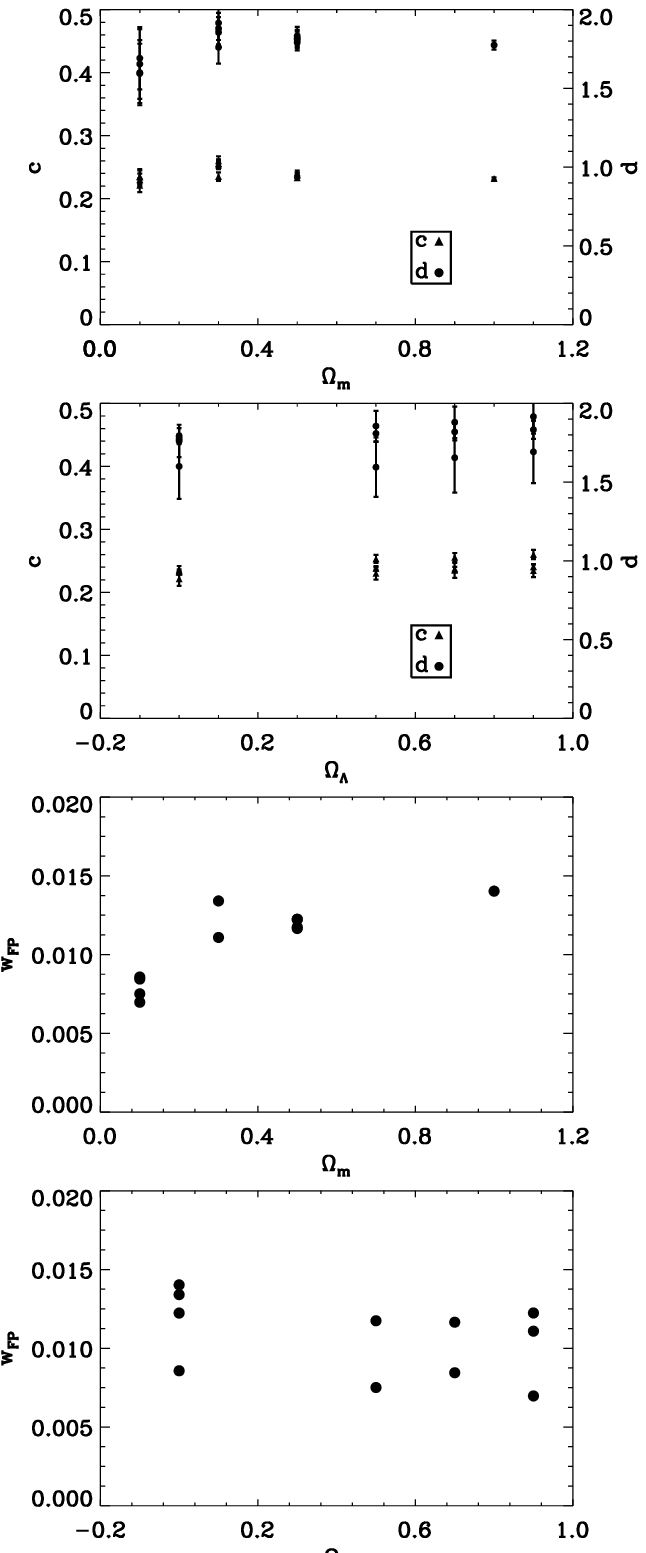}}
\caption{\small{Top panel: Fundamental Plane parameters $c$ (left axis, solid 
circles) and $d$ (righthand axis, solid triangles) as a function of 
$\Omega_{m}$. Second panel: Fundamental Plane parameters $c$ (left axis, solid 
circles) and $d$ (righthand axis, solid triangles) as a function of 
$\Omega_{\Lambda}$. Third panel: thickness $w_{fp}$ of the Fundamental Plane, 
i.e. rms scatter of the FP relation as a function of $\Omega_{m}$. Bottom 
panel: thickness $w_{fp}$ of the Fundamental Plane, ie. rms scatter of the FP 
relation as a function of $\Omega_{\Lambda}$.}}
\vspace{-0.5cm}
\label{fig:oms_fp}
\end{figure}

As can be seen in both Table~\ref{table:scaling_harm} and Fig. \ref{fig:sr_fp},
there is hardly any variation between the FP relations in the different 
cosmologies: they almost all coincide. This is certainly true concerning the FP
parameters $c$ and $d$. The two top panels of Fig.~\ref{fig:oms_fp} do confirm 
the impression that there is no systematic difference as a function of 
$\Omega_{m}$ and/or $\Omega_{\Lambda}$. This in itself is a strong argument 
against differences in the scaling relations parameters being due to a partial 
or incomplete level of virialization, as was claimed by \cite{adami98}.

One possible difference between the  Fundamental Plane in different cosmologies
may concern its thickness $w_{fp}$. Inspection of Fig.~\ref{fig:sr_fp} does 
suggest a marginally lower thickness of the FP for Universes with a low 
$\Omega_m\sim 0.1$.  There is no detectable effect at all with respect to the 
cosmological constant $\Omega_{\Lambda}$. We might understand a dependence on 
$\Omega_m$, or cosmological constant $\Omega_{\Lambda}$, in terms of the 
ongoing evolution of the cluster population. In low $\Omega_{m}$ Universes - 
and in high $\Omega_{\Lambda}$ universes all clusters formed at high redshift 
and have since had ample time to reach full virialization and hence tighten the
corresponding Fundamental Plane. In high $\Omega_{m}$ Universes, clusters would
still undergo a substantial levels of merging and accretion, both of which may 
affect the virial state of the cluster. Our computer experiments do not seem to
find any strong and significant dependence on overall cosmology. 

We investigate the relationship between the FP thickness and the dynamical 
state of the cluster in more detail in section \ref{sec5:mergacc}. 

Finally, we can try to relate the Fundamental Plane ($\mu$, $r_{h}$, 
$\sigma_v$) of our simulated cluster samples to the observationally measured 
($L$, $R_{e}$, $\sigma_v$) plane, e.g. $L\propto R^{1.19}\sigma^{0.91}$  found 
for the ENACS survey..  We can ask whether the difference can be ascribed 
solely to a mass dependent mass-to-light ratio $M/L$.  

\subsection{Scaling Relations for alternative Radius Definition}
Apart from the mean harmonic radius that we have used as a measure of halo size
in the previous sections, we have also assessed the viability of the scaling 
relations in case of alternative size definitions. In Table~\ref{table:halfvir}
we list the resulting parameters for the Kormendy relation and the Fundamental 
Plane in the case of using the half-mass radius $r_{half}$.

The parameters for the Kormendy relation hardly differ from the ones inferred 
on the basis of the mean harmonic radius.  However, the inferred Fundamental 
Plane plane parameters do differ significantly from the ones inferred above on 
the basis of the mean harmonic radius. The change in scaling parameter values
may be ascribed to the use of quantities that probe different aspect of the 
structure and dynamics of the halos. In an extreme situation, this might have 
disrupted the scaling relations. Our finding shows that the Kormendy relation 
still holds, while the FP relation still holds but in a slightly different 
guise. It may be an indication for our contention that halos do not form a 
perfectly a perfectly homologous population. Size measures sensitive to 
different aspects of the halos' internal mass distribution may then result in 
somewhat different scaling properties. In this respect, we agree with the 
conclusions of \cite{adami98} and \cite{lanzoni04}.

See section \ref{sec5:mergacc} and \ref{sec5:reconcile} for a discussion of the
relationship between the radii $r_{half}$ and $r_h$, where we show that it is a
consequence of the cluster building process.

\begin{figure*}
\centering
\includegraphics[width=0.90\textwidth]{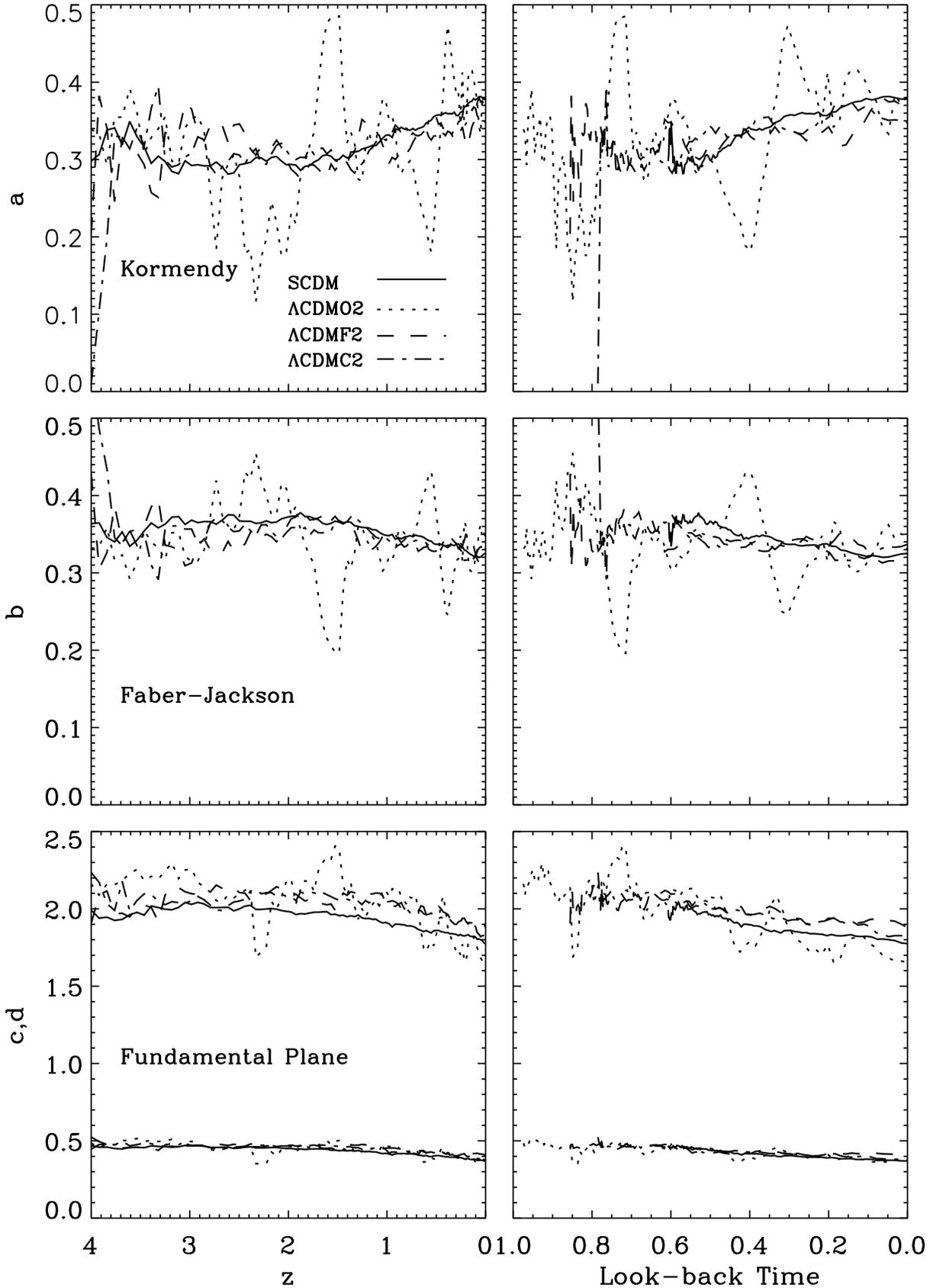}
\caption{{\small Evolution of the fitted scaling relation parameters as a 
function of redshift (left column) and as a function of cosmic look-back time 
(right column). Top: Kormendy parameter $a$. Center: Faber-Jackson parameter 
$b$. Bottom: FP parameters $c$ and $d$.}}
\label{fig:sr_z}
\end{figure*}

\begin{figure*}
\centering
\includegraphics[width=0.70\textwidth]{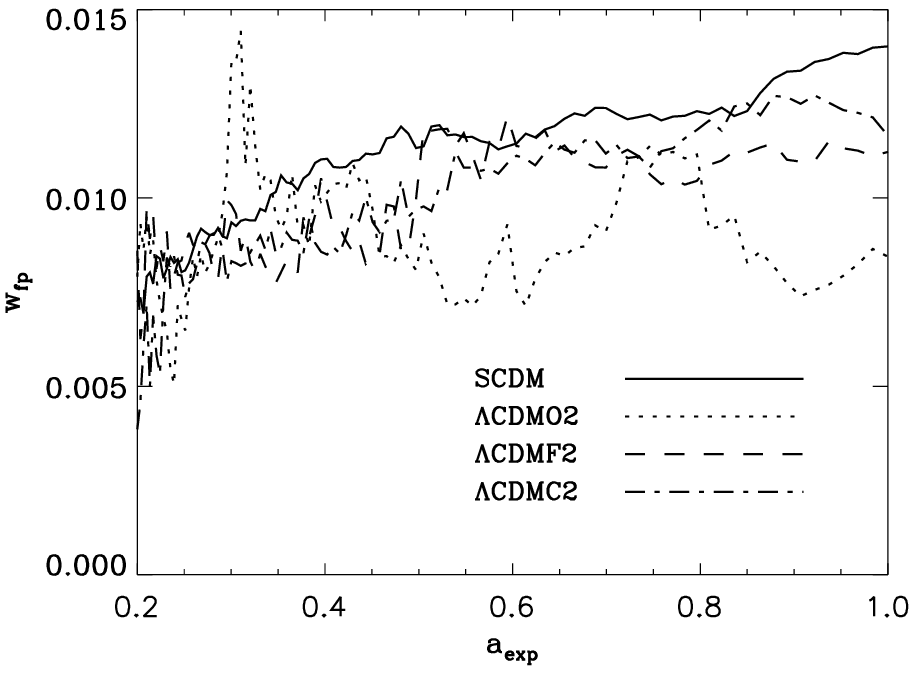}
\caption{\small{Evolution of the thickness of the Fundamental Plane for four
different cosmologies. Note the almost consistently tighter FP for the low
$\Omega_{m}$ Universe and the modest increase of FP thickness in the other 
cosmologies.  The SCDM, $\Lambda$CDMF2 and $\Lambda$CDMC2 models are very 
similar in their behaviour.  The more erratic behaviour of the $\Lambda$CDMO2 
may in part be due to the smaller sample size}}
\label{fig:spreadcosmo}
\end{figure*}

\section{Evolution of Scaling Relations}
\label{sec5:evol_scaling}
In the previous sections we have extensively studied the scaling relations at 
the current cosmic epoch $z=0$. We have also noted that there are differences 
between the scaling relation parameters that we find in our simulations and 
those for perfect virialized and homologous systems. This makes it interesting 
to trace the evolution of the different scaling relations.

In this section we investigate the evolution of the scaling relations as a 
function of redshift and as a function of cosmic look-back time.  While 
observers usually think in terms of redshift, it is important to appreciate 
that a given redshift corresponds to an entirely different dynamical epoch in 
different cosmologies. Given the same Hubble parameter, the age of the Universe
is a sensitive function of the cosmic density parameter $\Omega_{m}$ and even 
more so of the cosmological constant.  As for the latter, we have to realize 
that the change in cosmic time as a function of the cosmological constant is 
the most important influence of $\Lambda$. To give an appreciation of the 
differences in cosmic time for a given redshift in the different cosmologies, 
we refer to Table~\ref{table:ct}.

We have probed the scaling relations over a range of redshifts from $z=4$ to 
$z=0$ and over a range of cosmic look-back time going from $1$ to $10$ Gyr. The
evolution of the fitted scaling parameters as a function of redshift is shown 
in the left column of Fig. \ref{fig:sr_z}. The corresponding evolution as a 
function of cosmic look-back time can be found in the right hand column.  The 
Kormendy parameter $a$ is shown in the top panels, the Faber-Jackson parameter 
$b$ in the center panels and the FP parameters $c$ and $d$ in the bottom 
panels. Each different cosmology is represented by a different linestyle, 
listed in the insert at the top left hand frame.

\begin{figure*}
\centering
\includegraphics[width=0.80\textwidth]{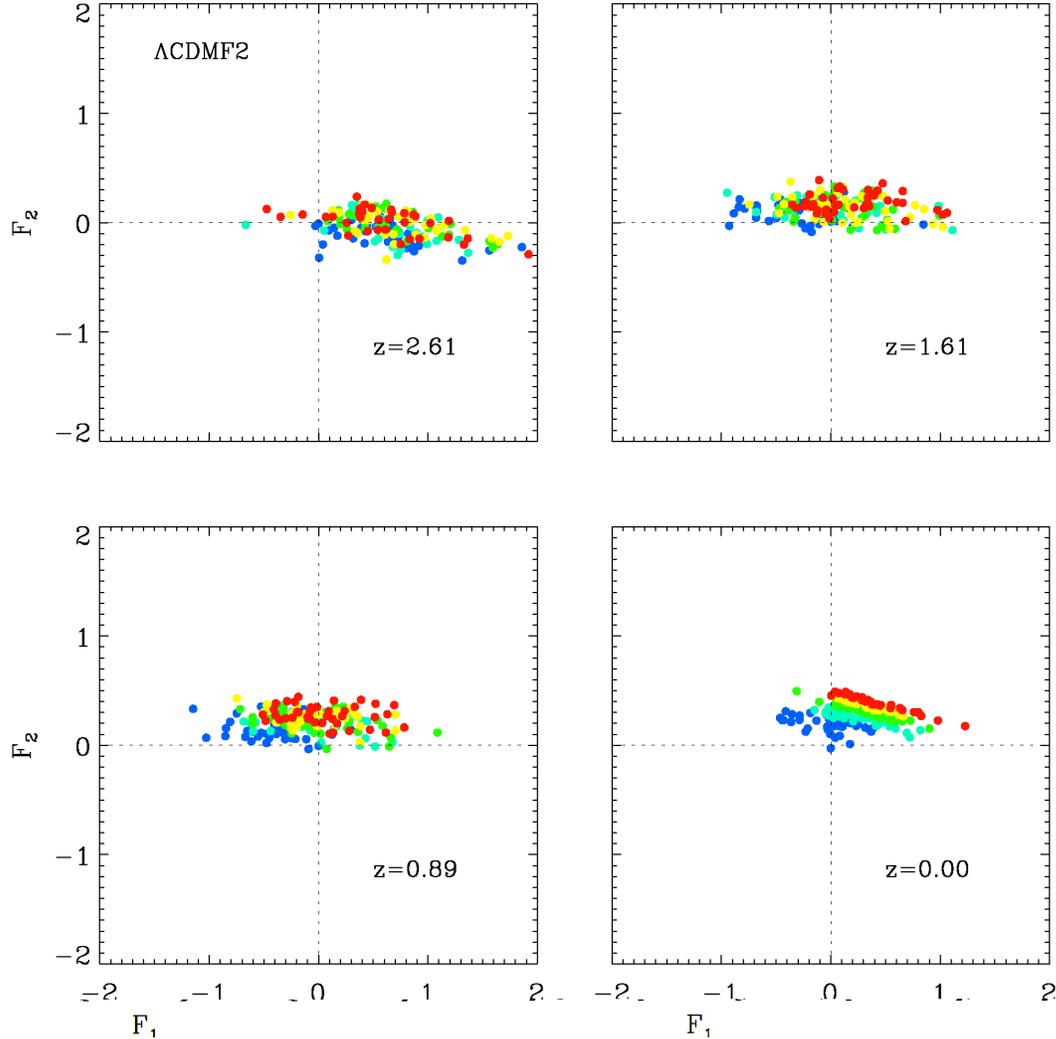}
\caption{{\small Shifting location of the cluster halo population within the 
Fundamental Plane. The depicted halo sample is the one in $\Lambda$CDMF2
cosmology, and is shown at four different redshifts: $z=2.61$ (top left panel),
$z=1.61$ (top right panel), $z=0.89$ (bottom left panel) and $z=0$ (bottom 
right panel). The abscissa and ordinate axis are arbitrarily chosen, mutually 
perpendicular, axes within the FP plane defined by
($\log{r_{h}},\mu,\log{\sigma_v}$) at $z=0$ (Eqn.~\ref{eq:fplcdm}).}}
\label{fig:Lambdaplanes}
\end{figure*}
\begin{figure*}
\centering
\includegraphics[width=0.80\textwidth]{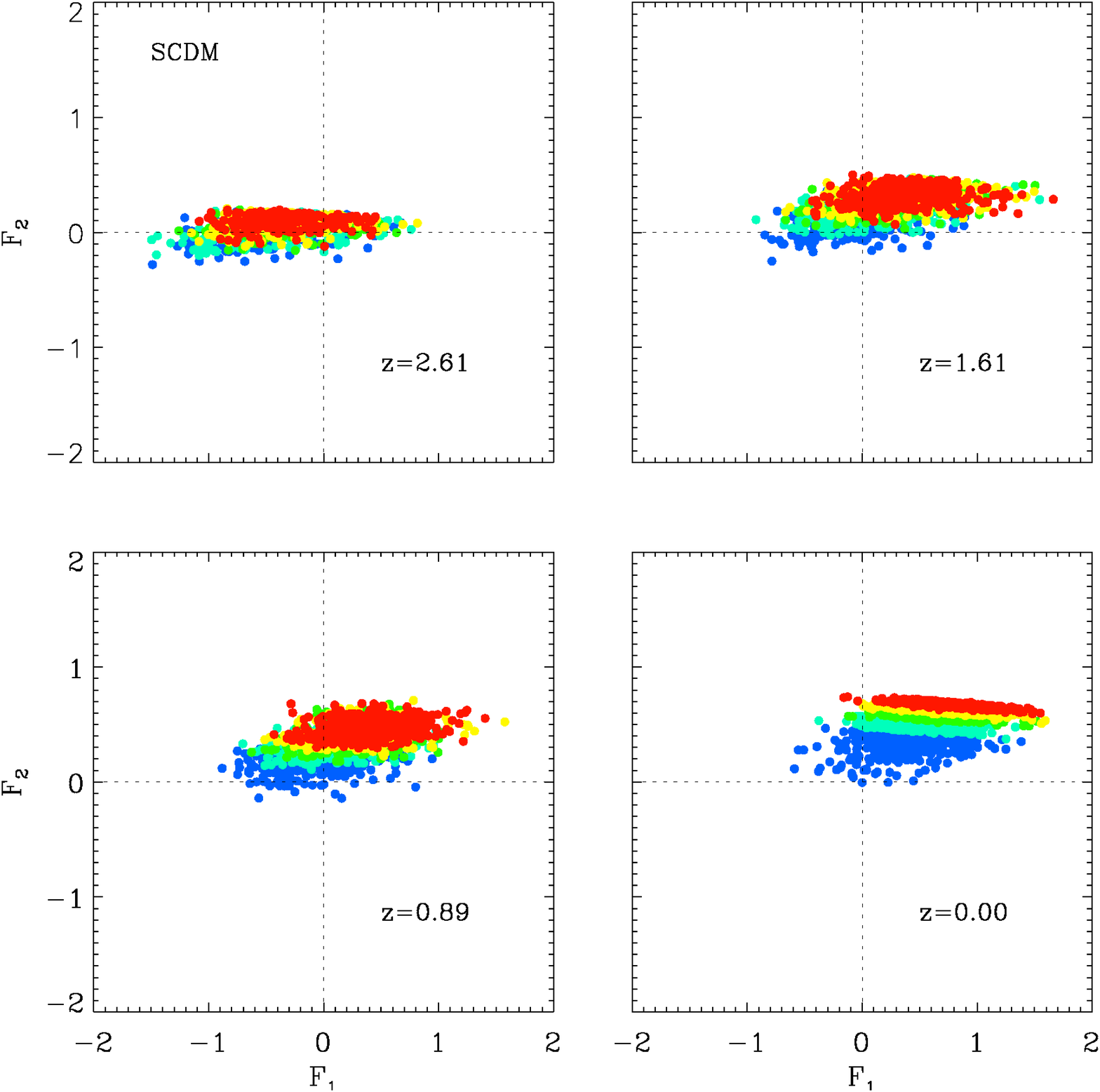}
\caption{{\small Shifting location of the cluster halo population within the 
Fundamental Plane. The depicted halo sample is the one in SCDM cosmology, and 
is shown at four different redshifts: $z=2.61$ (top left panel),
$z=1.61$ (top right panel), $z=0.89$ (bottom left panel) and $z=0$ (bottom 
right panel). The abscissa and ordinate axis are arbitrarily chosen, mutually 
perpendicular, axes within the FP plane defined by 
($\log{r_{h}},\mu,\log{\sigma_v}$) at $z=0$ (Eqn.~\ref{eq:fpscdm}).}}
\label{fig:SCDMplanes}
\end{figure*}

\subsection{Evolution of the Kormendy relation}
For all cosmologies the evolution of the Kormendy relation is marginal at best.
In the case of the low $\Omega_{m}$ $\Lambda$CDMO2 cosmology we can not discern
any significant change of the parameter $a$, (this may in part be due to the 
large uncertainties in the calculated parameter resulting from the low number 
of halos in this simulation). In the case of the other cosmologies we find no 
noticeable change of $a$ before a redshift $z\approx 2$, followed by a mild 
increase from $a \approx 0.3$ to $a\approx 0.38$ at $z \approx 0$. This is also
clearly visible when assessing the evolution in terms of cosmic time, as can be
seen in the top right panel.

\begin{figure*}
\centering
\includegraphics[width=0.6\textwidth]{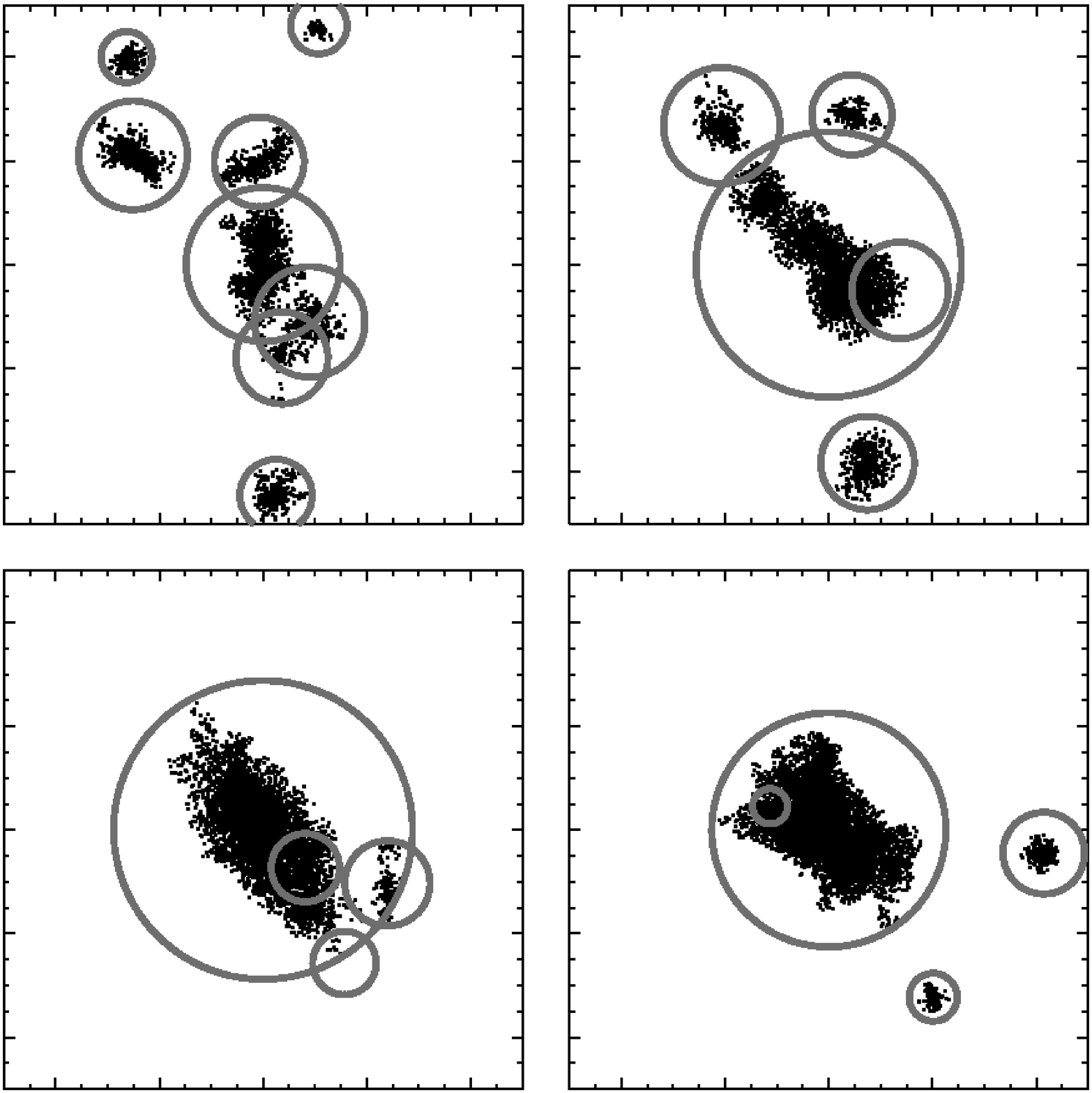}
\caption{Accreting vs. Merging Halo Evolution: the evolution of two different 
halos in the $\Lambda$CDMF2 cosmology. Each panel concerns a $5h^{-1}$ Mpc 
comoving size box centered on the core of the halo. The sequence runs from 
$z=2.61$ (top left panel), via $z=1.61$ (top right panel), to $z=0.89$ (bottom 
left panel) and finally the present epoch $z=0$ (bottom right panel). The 
circles indicate the location of HOP identified halos, with the size of the 
circle being proportional to the (virial) radius of the halo (overlapping 
circles are due to the projection of the corresponding spheres). Top four 
panels: a quiescently evolving accreting halo. Bottom four panels: a strongly 
hierarchically evolving merging halo.}
\includegraphics[width=0.6\textwidth]{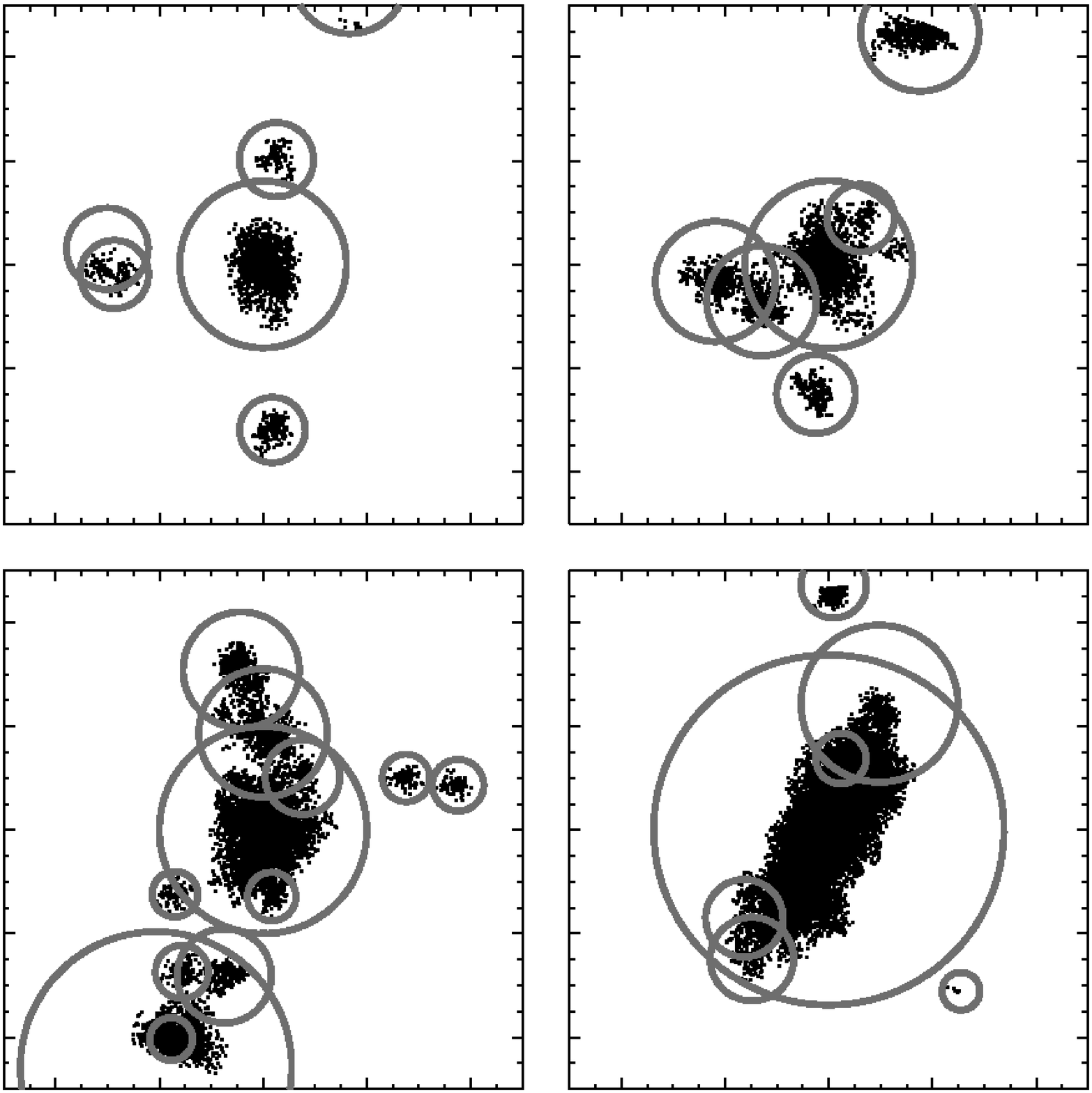}
\label{fig:merging_accretion}
\end{figure*}

\begin{figure*}
\centering
\includegraphics[width=0.80\textwidth]{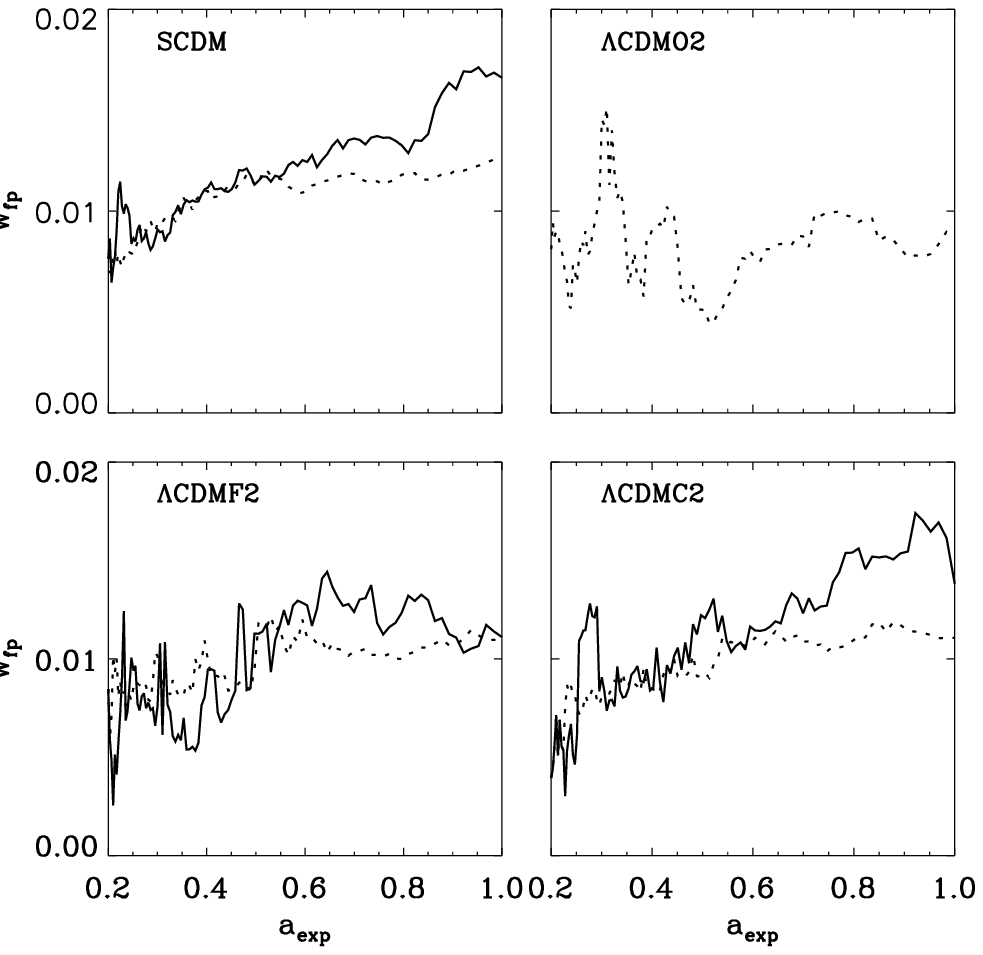}
\caption{\small{Thickness of the Fundamental Plane when considering accretion 
(dotted lines) or mergers (solid lines).}}
\label{fig:mer_acc_spread}
\end{figure*}

\subsection{Evolution of the Faber-Jackson relation}
Evolutionary trends for the Faber-Jackson relation are comparable to that seen 
in the Kormendy relation. No discernible trends are found in the open 
cosmology, while all of the other high density Universes do show a mild 
decrease from $b\approx 0.35$ at $z\approx 2$ to $b\approx 0.32$ at 
$z \approx 0$. When assessing in terms of cosmic time (center right panel), we 
observe a near uniform increase of $b$ over the last 8 Gyr. 

In most studied cosmologies, with the possible exception of the $\Lambda$CDMO2 
cosmology, we find a marginal trend of the Fundamental Plane parameter $c$ to 
decrease for $z <2$, more or less in the past $\sim$ 6-7 Gyr. At earlier epochs
such a trend is entirely absent. No significant evolution of the FP parameter 
$d$ can be observed in Fig.~\ref{fig:sr_z}.
 
\subsection{Evolution of the Fundamental Plane}
No significant evolution has been found for the Fundamental Plane parameters 
$c$ and $d$ (see Fig.~\ref{fig:sr_z}, lower panels). Evolution of the 
Fundamental Plane mainly concerns its thickness.

In Fig.~\ref{fig:spreadcosmo} we show the development of the FP thickness as a 
function of cosmic expansion factor $a_{exp}(t)=1/(1+z)$ for four cosmological 
models, and in Figs.~\ref{fig:SCDMplanes} and ~\ref{fig:Lambdaplanes} we show 
the evolution of the spread of points with the FP as a function of redshift in 
the $\Lambda$CDMF2 model.

We see a systematic increase of FP thickness over the whole cosmic evolution in
the case of the high $\Omega_{m}$ SCDM cosmology.  While we do see a rise of 
the FP thickness before $a_{exp}<0.5$ in the $\Lambda$CDMF2 and $\Lambda$CDMC2 
cosmologies, after that time the increase levels off and may even flatten 
completely. Note, however, that these simulations do not attain sufficient halo
mass resolution at higher redshifts: in these cosmologies halos still are low 
mass objects at these epochs. The one exceptional cosmology is that of the low 
$\Omega_{m}$ Universe $\Lambda$CDMO2. Except for a rather abrupt and sudden 
jump in FP thickness at $a_{exp}\sim0.3$, there is no noticeable change at 
later epochs. By $a_{exp}=0.3$ nearly all its clusters are in place and define 
a Fundamental Plane that does not undergo any further evolution. 
 
In summary, the trend seems to be for initial increase of the FP thickness 
followed by a convergence to a nearly constant value. The epoch of convergence 
is later for higher values of $\Omega_m$: while the thickness remains constant 
for the low $\Omega_m$ $\Lambda$CDMO2 cosmology, it involves a slow but 
continuous increase in the SCDM cosmology. 

On the basis of their study of galaxy merging, \cite{nipoti03} argued that the 
disposition of galaxies in the Fundamental Plane is not simply a realization of
the virial theorem, but contains additional information on galaxy structure and
dynamics.  This should be reflected in the location of the halo population 
within the Fundamental Plane. 

Figs.~\ref{fig:Lambdaplanes} and ~\ref{fig:SCDMplanes} show how the location 
of the clusters within the plane shifts as time proceeds. The color scheme is 
the same as for Fig.~\ref{fig:radiusrel}. Fig.~\ref{fig:Lambdaplanes} shows the
location of the clusters in the $\Lambda$CDMF2 cosmology in the Fundamental 
Plane inferred for the current epoch, ie. at redshift $z=0$, 
\begin{equation}
\log{r_h}\,=\,0.41\,\mu\,+\,1.88\,\log{\sigma_v}\,+C_{fp,L}\,.
\label{eq:fplcdm}
\end{equation}
To locate their position within the Fundamental Plane, we use the (artificial) 
coordinates $F_1$ and $F_2$ of the halo points with respect to two mutually 
perpendicular normalized vectors in the Fundamental Plane at $z=0$, wrt. the
coordinate system defined by the FP quantities ($\log{r_h},\mu,\log{\sigma_v}$)
(note that $F_1$ and $F_2$ do not have a specific physical significance). From 
the panels in the figure we see that the evolution of halos involves a gradual 
shift along an almost universal Fundamental Plane. It also shows that the halo 
population seems to evolve from a more scattered and somewhat looser one into a
tightly  elongated point cloud at the current epoch, providing interesting 
clues towards understanding the cluster virialization process. 

In the same vein, Fig.~\ref{fig:SCDMplanes} follows the changing location of 
clusters in the SCDM cosmology in the corresponding Fundamental Plane at $z=0$,
\begin{equation}
\log{r_h}\,=\,0.37\,\mu\,+\,1.78\,\log{\sigma_v}\,+C_{fp,S}\,.
\label{eq:fpscdm}
\end{equation}
Similar to the $\Lambda$CDMF2 cosmology, we find that the cluster point cloud 
appears to assume a clearer mass stratification as time proceeds. While the 
population of clusters in the SCDM cosmology also appears to shift its location
along the Fundamental Plane as it evolves, we do not find a trend towards a 
more tightly point cloud that we see in the LCDMF2 cosmology. We will 
investigate these evolutionary trends in more detail in an upcoming study, we 
have found indications for a possible influence of the different cluster halo 
merging histories in SCDM and $\Lambda$CDMF2 in explaining the different 
behaviour of the cluster point clouds in the Fundamental Plane. 

\begin{figure}
\centering
\includegraphics[width=0.45\textwidth]{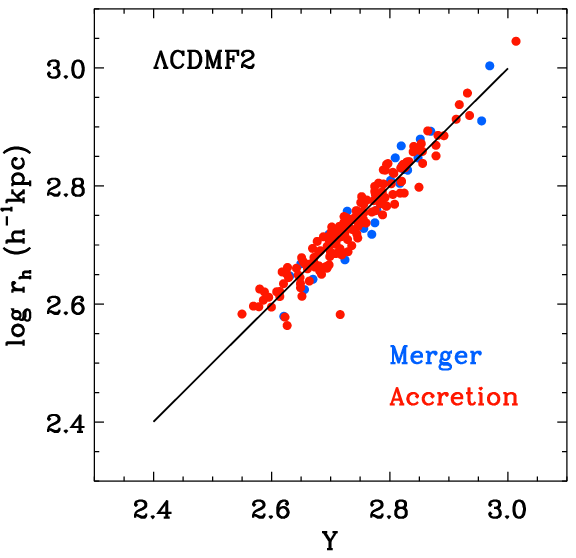}
\caption{\small{Comparison between the Fundamental Plane in the $Lambda$CDMF2 
cosmology for clusters that underwent a major merger (blue dots) and clusters 
that followed a more quiescent accretion history (red dots). The plot depicts 
the relation between harmonic radius $r_{h}$ and the quantity 
$Y=c\mu + d \log\sigma + C_{fp}$, in which $c$ and $d$ are the FP scaling 
parameters.}}
\label{fig:fp_merg_acc}
\end{figure}

\section{Merging and accretion dependence}
\label{sec5:mergacc}
Figures Fig. \ref{fig:sr_z} and Fig.~\ref{fig:spreadcosmo} show that the 
evolution of the parameters defining these relationships is very erratic.  This
testifies to the fact that in hierarchical structure formation scenarios the 
formation and evolution of halos is hardly a quiescent and steadily progressing
affair. Rather, halos grow in mass by steady accretion of matter from its 
surrounding as well through the merging with massive peers. Even the accretion 
is not a continuous and spherically symmetric process: most matter flows in in 
a strongly anisotropic fashion through filamentary extensions into the 
neighboring large scale matter distribution. As a result, we can expect that 
many halos will not have settled into a perfect virial state. This will 
certainly be the case for halos that recently suffered a major merger with one 
or more neighboring clumps. 

The detailed accretion and merging history is a function of the underlying 
cosmology. Low density cosmologies or cosmologies with a high cosmological 
constant will have frozen their structure formation at early epochs. The halos 
that had formed by the time of that transition will have had ample time to 
settle into a perfect virialized object. Also, there is a dependence on the 
power spectrum of the corresponding structure formation scenario. Power spectra
with a slope $n<-1.5$ (at cluster scales) will imply a more homologous collapse
of the cluster sized clumps, less marked by an incessant bombardment by smaller
clumps. It may be clear that a more violent life history of a halo will usually
be reflected in a substantial deviation from a perfect virial state.

In order to investigate the implications of a difference in accretion or 
merging history of halos, we have split the samples of cluster halos in each 
of our cosmologies into a \emph{merging sample} and a \emph{accretion sample}. 
Possible differences in their virial state should be reflected in the quality 
of the scaling relations, in particular that of the thickness of the 
Fundamental Plane.  

The \emph{merger sample} consists of those halos that suffered a merger with 
another halo that contained at least 30\% of its mass. 
Fig.~\ref{fig:merging_accretion}  shows two examples of halos in the 
$\Lambda$CDMF2 cosmology. The top sequence of 4 panels shows the evolution of a
quiesencently evolving {\it accretion halo}, by means of the particle 
distribution in a $5h^{-1}$ Mpc box (comoving size) around the cluster core, at
z=2.61, z=1.61, z=0.89 and $z=0.00$. The circles indicate the location of the 
HOP identified halos, with the size of the circle proportional to the radius of
the halo (note that the overlap of circles is due to projection of the halo 
spheres). The lower group of 4 panels shows the particle distribution at the 
same redshifts for a halo belonging to the \emph{merging sample}. Its gradual 
hierarchical buildup is directly visible as the the continuous infall of clumps
at each timestep. 

In  Fig.~\ref{fig:mer_acc_spread} we show the evolution of the thickness of the
Fundamental Plane for each of the two samples in the four indicated 
cosmologies. Note that our simulations do not have sufficient resolution for 
reconstructing the precise merging or accretion history before 
$a_{exp}=0.3-0.4$, so that we may not draw conclusions on the rise of the FP 
thickness up to that epoch. Also, in the case of the $\Lambda$CDMO2 scenario we
do not have enough cluster halos to be able to detect any systematic 
differences between the merging and accreting halos.

In the more recent history we do find some significant differences between 
merging and accretion-only halos in various cosmologies, in particular the ones
with a high $\Omega_m$. There does not seem to be a systematic difference 
between these groups in the $\Lambda$CDMF2 cosmology. The total absence of any 
difference between the Fundamental Plane of merging and accreting cluster halos
at present (Fig.~\ref{fig:fp_merg_acc}), is the outcome of an evolutionary 
history without any significant differences between the two subsamples 
(Fig.~\ref{fig:mer_acc_spread}, lower lefthand panel). 

\begin{figure}
\centering
\vskip -0.5truecm
\includegraphics[width=0.45\textwidth]{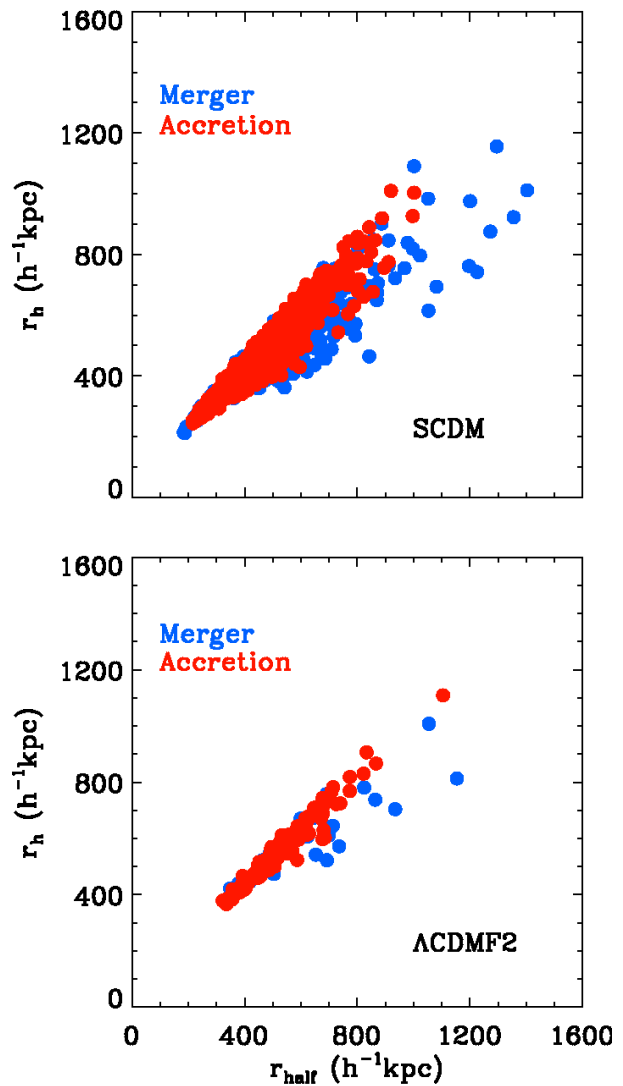}
\caption{\small{Comparison between the mean harmonic radius and the half-mass 
radius of the halos in the SCDM (top) and $\Lambda$CDMF2 (bottom) scenarios. 
The cluster samples are split into the clusters that underwent a major merger 
(blue dots) and the ones that accreted matter in a more quiescent fashion (red 
dots). See text for further explanation.}}
\label{fig:mergaccrad}
\end{figure}
The story is quite different for the $\Lambda$CDMC2 and SCDM cosmology. While 
the cluster halos that undergo a major merger do reveal a constantly growing FP
thickness, their accretion-only clusters do not display such a systematic 
increase. Instead, their FP thickness remains lower and levels off. In other 
words, accretion halos (dotted lines) do on average display a tighter FP 
relation. This is particularly true at the current epoch. Apparently, the 
absence of violent mass gain in the case of accretion halos implies that they 
have more time to relax and virialize. This, in turn, is reflected in a 
thickness of the Fundamental Plane which does not evolve any further. 
Interestingly, it is also reflected in the radii of the halos 
(Fig.~\ref{fig:mergaccrad}): while the harmonic radius and half-mass radius of 
accreting cluster halos are mostly in accordance with each other, though with a
larger spread than in the case of $\Lambda$CDMF2 clusters (lower panel), the 
SCDM halos that underwent major mergers do appear to be responsible for the 
substantial differences between the harmonic and half-mass radii that we see in
Fig.~\ref{fig:radiusrel}. From this we conclude that the accretion history is a
major factor in determining the character of the Fundamental Plane, via the 
impact of mergers on the mass distribution within halos and hence their radii. 
We discuss this in more detail in the next section~\ref{sec5:reconcile}.

The implications of this finding might be far-reaching. Given the remarkable 
robustness and stability of the Fundamental Plane, any deviation of individual 
clusters from the FP may be a direct reflection of its recent dynamical 
evolution. This would be true if the thickness of the plane would be entirely 
due to the merger history of the clusters. It is certainly a viable implication
of our conclusion that the Fundamental Plane's definition -- the average plane 
of a large sample of clusters -- is nearly unassailable while we find strong 
fluctuations and deviations from the average FP in small samples of actively 
evolving clusters. 

In practice, it might mean that one could take samples of clusters in different
redshift bands and reliably average them in each band to use the resulting 
Fundamental Plane to study redshift evolution of observed samples. It would 
also mean that within each redshift band you know which ones have had active 
lives.

\section{Reconciling the simulations with the virial theorem}
\label{sec5:reconcile}
\subsection{The Virial Theorem}
The Fundamental Plane is a direct reflection of the virial theorem which, under
particular assumptions, relates the averaged velocity dispersion and radius of 
a system directly to its mass.  All ``virialised'' objects will lie on a plane 
defined in the space of those three variables.  There is not even any freedom 
in the parameters for that plane: its slope and location are fixed for all 
virialised objects.

There are complications when assigning data to a Fundamental Plane.  Firstly, 
in its simplest form, the virial theorem assumes that the virialised objects 
are isolated spherical systems and, importantly, that they are stationary.  The
systems we study are not spherical and they are certainly not stationary: they 
are generally in a state of dynamical evolution.  The possible exception to 
this might be the largest most isolated systems.  Secondly, observed data does 
not have direct knowledge of the system mass except through interpreting the 
light that is observed.  The universality of the Fundamental Plane allows us to
turn the problem around and determine the dependence of light on mass in order 
that systems should fit on the Fundamental Plane. The simplest approach to this
is to assume that the mass to light ratio in the observed waveband is directly 
related to mass.

There are further issues.  For example, what do we mean when we refer to 
``averages'' of quantities?  Using a different averaging process yields a 
different Fundamental Plane.  There is also the fact that astrophysical systems
are observed only in projection.

Having said that, we can express the Virial Theorem in terms of the variable we
have used here to describe the Fundamental Plane.  With the notation that a 
virialised system of mass $M$ has a velocity dispersion $V$, half mass radius 
$r_{half}$ and harmonic  radius $r_h$ we have, up to normalising constants:
\begin{equation}
V^2 = \frac{M}{r_h^2}, \quad \Sigma=\frac{M}{r_{half}^2}
\end{equation}
where $\Sigma$ is the projected (surface) mass density. Eliminating $M$ from 
these and taking logs yields an expression for the Fundamental Plane:
\begin{equation}
\log r_h = \left( \frac{r_h}{r_{half}}\right) + 2 \log V + 0.4\mu
\label{eq:virial}
\end{equation}
where we have transformed the surface mass density $\Sigma$ into logarithmic 
astro-units via 
\begin{equation}
\mu = -2.5 \log \Sigma
\end{equation}
We have explicitly written equation (\ref{eq:virial}) in such a way as to 
expose the different roles of harmonic and half-mass (geometric) radii. The 
relationship between these radii in our models is illustrated in 
Fig.~\ref{fig:radiusrel} and in Fig.~\ref{fig:mergaccrad}, the latter 
differentiating between merging and quiescently accreting halos. 
 
\subsection{Renormalising the FP Simulations}
It is important to understand why the coefficients of the model Fundamental 
Plane might differ from the expectations based on the use of the virial 
theorem.  Luminosity is not involved here so we cannot appeal to a varying 
mass-to-light ratio.  Moreover, the model Fundamental Plane is well defined and
so we cannot say that this is merely a question of fitting.

The are at least two possible sources for this systematic difference between 
the model and the virial theorem.  The first is to blame the HOP technique and 
assert that it systematically underestimates the cluster masses.  The second is
to say that the internal cluster properties (like velocity distribution) vary 
systematically with mass an so the normalisation of the virial plane is mass 
dependent.

Either way, we shall model in a mass dependency and consider this in relation 
to the HOP technique.  The process for the variable virial normalisation is 
analogous.

The samples of clusters derived from these simulations are all based on the HOP
technique.  There may well be a systematic bias in the assignment of particles 
to clusters (see section \ref{sec:haloid}). As a consequence, the radii and 
velocity dispersion derived for a HOP selected cluster will also be biased.  
Clearly the bias will be more significant for smaller systems. 

In this subsection we seek to account for systemic effects of using HOP for 
identifying cluster membership, and derive a renormalisation procedure taking 
account of this and matching the dataset to the expected virial theorem 
Fundamental Plane (equation~\ref{eq:virial}).   

The easiest way to model this bias is to assume that the model-based estimate 
(biased) for the mass, $M$, is related to the actual mass $\cal{M}$ by a simple
scaling relationship
\begin{equation}
\frac{\cal{M}}{M} \propto {\cal{M}} ^{\frac{\alpha}{1+\alpha}}
\end{equation}
for some exponent $\alpha$.  The virial expression for the mass then becomes
\begin{equation}
V^2 = \left(\frac{\cal{M}}{M} \right)\frac{M}{r_h^2}= M^{-\alpha} \frac{M}{r_h^2}
\end{equation} 
where the right hand side now refers to quantities derived from the model. We 
can eliminate $M$ from this in terms of the model surface mass density 
$\Sigma = M/{r_h^2}$ to give
\begin{equation}
r_h= \left(\frac{r_h}{r_{{half}}} \right)^{\frac{2}{1+\alpha}} V^{2\left(\frac{1-\alpha}{1+\alpha} \right)} \Sigma^{-\frac{1}{1+\alpha}}
\end{equation} 
Taking logs and using $\mu = -2.5log \Sigma$ finally yields
\begin{equation}
\log r_h = \frac{2}{1+\alpha} 
\log \frac{r_h}{r_{{half}}} + 
2 \frac{1-\alpha}{1+\alpha} \log V - 
\frac{0.4}{1+\alpha} \mu
\label{eq:biasfit}
\end{equation} 
which is the expression for the Fundamental Plane in terms of the (biased) 
model derived quantities.  This should be compared with equation 
(\ref{eq:virial}): we see how the bias modelled by $\alpha$ affects the 
position and slope of the virial Fundamental Plane.

The procedure now, for each simulation, is to select a value of $\alpha$ that 
makes the coefficient of $\log V$ in equation (\ref{eq:biasfit}) equal to the 
virial value $2$.  That $\alpha$ then allows a calculation of the coefficient 
of $\mu$ that can be compared with the value derived from the simulation.  The 
results for a selection of models is shown in Table \ref{table:biasfit}.

\begin {table}
  \begin {center}
    \begin {tabular}{||c|c|c|c|c||}
      \hline
      Model & d   & $\alpha$ & $c$ & $c$(model) \\
      \hline
      \hline
      SCDM   & $1.78 \pm 0.01  $ & $0.058 \pm 0.003 $ & $0.38$ & $0.37 \pm 0.031 $ \\
      LCMDF1 & $1.69 \pm 0.084 $ & $0.084 \pm 0.024 $ & $0.37$ & $0.38 \pm 0.016 $ \\
      LCDMF2 & $1.88 \pm 0.04  $ & $0.031 \pm 0.011 $ & $0.39$ & $0.41 \pm 0.011 $ \\
      LCDMF3 & $1.81 \pm 0.02  $ & $0.050 \pm 0.014 $ & $0.38$ & $0.38 \pm 0.056 $ \\
      OCDM01 & $1.60 \pm 0.083 $ & $0.111 \pm 0.024 $ & $0.36$ & $0.35 \pm 0.017 $ \\
      \hline
    \end {tabular}
    \caption{\small{Fitting biased models to ideal virial Fundamental Plane.  
The parameter $\alpha$ emulates the limitations of the HOP group finder.}}
    \label{table:biasfit}
  \end {center}
\end {table}

The conclusion to be drawn from this is that, for each model, there is indeed a
value of the $\alpha$ parameter that reproduces the Fundamental Plane fits for 
the models.  

\subsection{Observed cluster FP}
The best available data set is the ENACS data of \citep{adami98}. 
Equation~(\ref{eq:enacs}) describing that Fundamental Plane, in the current 
notation, reduces to
\begin{equation}
\log{R_e}\, =\,(0.49 \pm 0.05) \mu\,+\,(1.12 \pm 0.11) \log{\sigma_{v}}
\end{equation}
There is considerable uncertainty in this relationship: the coefficient of 
$\log \sigma$ is quite far from the ideal $2.0$ and the coefficient of $\mu$ is
higher than the nominal $0.4$.

The usual way to reconcile this with the virial Fundamental Plane is to argue 
that the mass to light ratio of the cluster sample is mass dependent:
\begin{equation}
\frac{M}{L} \propto M^\beta
\end{equation}
Using an argument that parallels the derivation of equation (\ref{eq:biasfit}),
the Fundamental Plane expressed in terms of velocity and surface mass density is
\begin{equation}
\log r_h = \frac{2}{1+\beta} 
\log \frac{r_h}{r_{{half}}} + 
2 \frac{1-\beta}{1+\beta} \log V - 
\frac{0.4}{1+\beta} \mu
\label{eq:enacsfit}
\end{equation} 
The data give $\beta=0.28 \pm 0.19$ which gives rise to $c = 0.31 \pm 0.02 $, a
long way from the data-derived $0.49$. It is clearly not possible to reconcile 
the ENACS data with the virial theorem Fundamental Plane, let alone the 
numerical simulations.  

\section{Conclusions and Discussion}
\label{sec5:conclusions}
We have studied three structural scaling relations of galaxy clusters in 
thirteen cosmological models. These relations are the Kormendy relation, the 
Faber-Jackson relation and the Fundamental Plane. Their validity and behavior 
in the different cosmological models should provide information on the general 
virial status of the cluster halo population. The cosmological models that we 
studied involved a set of open, flat and closed Universes with a range of 
matter density parameter $\Omega_{m}$ and cosmological constant 
$\Omega_{\Lambda}$. 

The cluster samples are obtained from a set of $N$-body simulations in each of 
the cosmologies. These simulations concerns a box of 200$h^{-1}$Mpc with 
256$^{3}$ dark matter particles. The initial conditions were set up such that 
the phases of the Fourier components of the primordial density field are the 
same for all simulations. In this way, we have simulations of a comparable 
morphological character: the same objects can be recognized in each of the 
different simulations (be it at a different stage of development).

After running the simulations from $z=4$ to the current epoch using the GADGET2
code, we used HOP to identify the cluster halos. We investigated whether each 
halo population obeyed a mass-radius relation akin to the Kormendy relation, a 
mass-velocity dispersion relation similar to the Faber-Jackson relation and a 
two parameter family between mass, radius and velocity dispersion that 
resembles a Fundamental Plane relation. We studied the dependence of the 
obtained scaling parameters as a function of the underlying cosmology and 
investigated their evolution in time.

\noindent Our results can be summarized as follows:

\begin{itemize}

\item In each cosmological model we do recover Kormendy, Faber-Jackson and 
Fundamental Plane relations for the population of cluster halos.  This is a 
strong indication that the halos are in a virialized state, as expected in 
hierarchical clustering scenarios.

\item There are significant differences between the measured parameters of the 
various scaling relations and those seen in the observational data. Our fit for
the FP in the $\Lambda$CMDF2 model is
\begin{equation}
\log r_h = 0.41 \mu + 1.86 \log \sigma + const.
\end{equation}
This can be reconciled with the expectation from the virial theorem, but not 
with the ENACS Fundamental Plane.

\item We do not find any significant dependence of the parameters $a$ and $b$ 
of the Kormendy and Faber-Jackson relations on the value of $\Omega_{m}$. There
is also no indication for any influence of $\Omega_{\Lambda}$ on the scaling 
relations. 

\item While the FP parameters $c$ and $d$ are not dependent on $\Omega_m$ and 
$\Omega_{\Lambda}$, there is a slight suggestion that the Fundamental Plane 
would have a lower thickness for low $\Omega_m\sim 0.1$ cosmologies. 

\item With the exception of low $\Omega_{m}$ Universes, we find a mild increase
of the Kormendy parameter $a$ and a mild decrease of the Faber-Jackson 
parameter $b$ from $z=1$ to the present epoch. From $z=4$ to $z=1$ we did not 
find any discernable evolution. 

\item While the Fundamental Plane parameters $c$ and $d$ do in general not show
a significant evolution, the higher $\Omega_m$ cosmologies do involve a slight 
decrease of FP parameter $d$ during most recent epochs ($z < 2$).  

\item The thickness of the Fundamental Plane does evolve significantly, with an
initial increase followed by a convergence to a more ore less constant value. 
The convergence epoch is later for higher density cosmologies. This probably 
reflects the gradually virializing tendency of the cluster population.

\item Given our expectation that there is a difference in virial state between 
quiescently accreting clusters and those experiencing massive mergers, we have 
investigated the evolution of the Fundamental Plane thickness for samples of 
merging clusters and samples of accreting clusters. We find that accreting 
clusters at recent epochs do appear to be better virialized than the merging 
population and that the FP thickness is smaller in the former.

\item We find that for all investigated cosmologies the Fundamental Plane is 
remarkably stable, despite the enormous evolution of the individual systems. 
The only significant evolution, that of its thickness, might be due in a large 
part to the importance of merging of individual systems. 

\item If indeed the thickness of the Fundamental Plane might be entirely due to
the merger history of the cluster halos, the distance of an individual cluster 
to the Fundamental Plane would be a direct reflection of the cluster history. 

\item
We see direct evidence that major mergers have effected the relationship 
between the galaxy haloes in the cluster in that the relationship between the 
half-mass and harmonic radii is disturbed.  Nonetheless, the evidence from the 
models tells us that this does not affect the slope of the Fundamental Plane: 
clusters that have undergone major mergers lie in the same place as those that 
have grown by steady accretion. 
\end{itemize}

Finally, what is desperately needed is better data on the cluster fundamental 
plane.  We might speculate that the distance of a cluster from the plane 
defined by the data somehow reflects the evolution of the cluster, but we will 
not get evidence for the hypotheses derived from numerical experiment until 
there is more high quality data.

\section*{Acknowledgements}
P.A. gratefully acknowledges support by NOVA. RvdW is grateful for the support 
and great hospitality of KIAS during the completion of this manuscript. In 
addition,  BJ gratefully acknowledges the hospitality of the Kapteyn 
Astronomical Institute in Groningen, and to his collaborators for their 
remarkable patience in getting various parts of this paper completed.

\bsp

\label{lastpage}

\end{document}